\documentclass[preprint,journal]{./vgtc}       % preprint (journal style)

\ifpdf%                                % if we use pdflatex
  \pdfoutput=1\relax                   % create PDFs from pdfLaTeX
  \usepackage{graphicx}                % allow us to embed graphics files
  \DeclareGraphicsExtensions{.pdf,.png,.jpg,.jpeg} % for pdflatex we expect .pdf, .png, or .jpg files
\else%                                 % else we use pure latex
  \usepackage{graphicx}                % allow us to embed graphics files
  \DeclareGraphicsExtensions{.eps}     % for pure latex we expect eps files
\fi%

\graphicspath{{figures/}{pictures/}{images/}{./}} % where to search for the images

\usepackage{microtype}                 % use micro-typography (slightly more compact, better to read)
\PassOptionsToPackage{warn}{textcomp}  % to address font issues with \textrightarrow
\usepackage{textcomp}                  % use better special symbols
\usepackage{mathptmx}                  % use matching math font
\usepackage{times}                     % we use Times as the main font
\usepackage{cite}                      % needed to automatically sort the references
\usepackage{tabu}                      % only used for the table example
\usepackage{booktabs}                  % only used for the table example
\usepackage{xcolor}
\usepackage{paralist}
\usepackage[hyphens]{url}
\usepackage{amsmath}
\usepackage{amssymb}

\usepackage{algorithm}
\usepackage[noend]{algpseudocode}
\usepackage{varwidth}
\algdef{SE}[DOWHILE]{Do}{doWhile}{\algorithmicdo}[1]{\algorithmicwhile\ #1}%
\renewcommand*\Call[2]{\textproc{#1}(#2)}

\usepackage{amsmath}
\DeclareMathOperator*{\argmax}{arg\,max}
\DeclareMathOperator*{\argmin}{arg\,min}

\usepackage{siunitx}

\onlineid{1235}

\vgtccategory{Research}
\vgtcpapertype{application/design study}

\title{Towards Better Bus Networks: A Visual Analytics Approach}

\author{Di Weng, Chengbo Zheng, Zikun Deng, Mingze Ma, Jie Bao, Yu Zheng, Mingliang Xu, Yingcai Wu}
\authorfooter{
\item
 D. Weng, Z. Deng, and Y. Wu are with State Key Lab of CAD\&CG, Zhejiang University, Hangzhou, China and Zhejiang Lab, Hangzhou, China. E-mail: \{dweng, zikun\_rain, ycwu\}@zju.edu.cn. Y. Wu is the corresponding author.
\item
 C. Zheng and M. Ma are with Zhejiang Lab, Hangzhou, China. Email: chbozheng@gmail.com, mamzaug@foxmail.com.
\item
 Jie Bao and Yu Zheng are with JD Intelligent Cities Research, Beijing, China and JD Intelligent Cities Business Unit, JD Digits, Beijing, China. E-mail: baojie@jd.com, msyuzheng@outlook.com.
\item
 Mingliang Xu is with School of Information Engineering, Zhengzhou University, Zhengzhou, China and Henan Institute of Advanced Technology, Zhengzhou University, Zhengzhou, China. E-mail: iexumingliang@zzu.edu.cn.
}

\shortauthortitle{Biv \MakeLowercase{\textit{et al.}}: Global Illumination for Fun and Profit}
\abstract{Bus routes are typically updated every 3--5 years to meet constantly changing travel demands. However, identifying deficient bus routes and finding their optimal replacements remain challenging due to the difficulties in analyzing a complex bus network and the large solution space comprising alternative routes. Most of the automated approaches cannot produce satisfactory results in real-world settings without laborious inspection and evaluation of the candidates. The limitations observed in these approaches motivate us to collaborate with domain experts and propose a visual analytics solution for the performance analysis and incremental planning of bus routes based on an existing bus network. Developing such a solution involves three major challenges, namely, a) the in-depth analysis of complex bus route networks, b) the interactive generation of improved route candidates, and c) the effective evaluation of alternative bus routes. For challenge a, we employ an overview-to-detail approach by dividing the analysis of a complex bus network into three levels to facilitate the efficient identification of deficient routes. For challenge b, we improve a route generation model and interpret the performance of the generation with tailored visualizations. For challenge c, we incorporate a conflict resolution strategy in the progressive decision-making process to assist users in evaluating the alternative routes and finding the most optimal one. The proposed system is evaluated with two usage scenarios based on real-world data and received positive feedback from the experts.%
} % end of abstract


\CCScatlist{ % not used in journal version
 \CCScat{K.6.1}{Management of Computing and Information Systems}%
{Project and People Management}{Life Cycle};
 \CCScat{K.7.m}{The Computing Profession}{Miscellaneous}{Ethics}
}

\teaser{
  \centering
  \includegraphics[width=\linewidth]{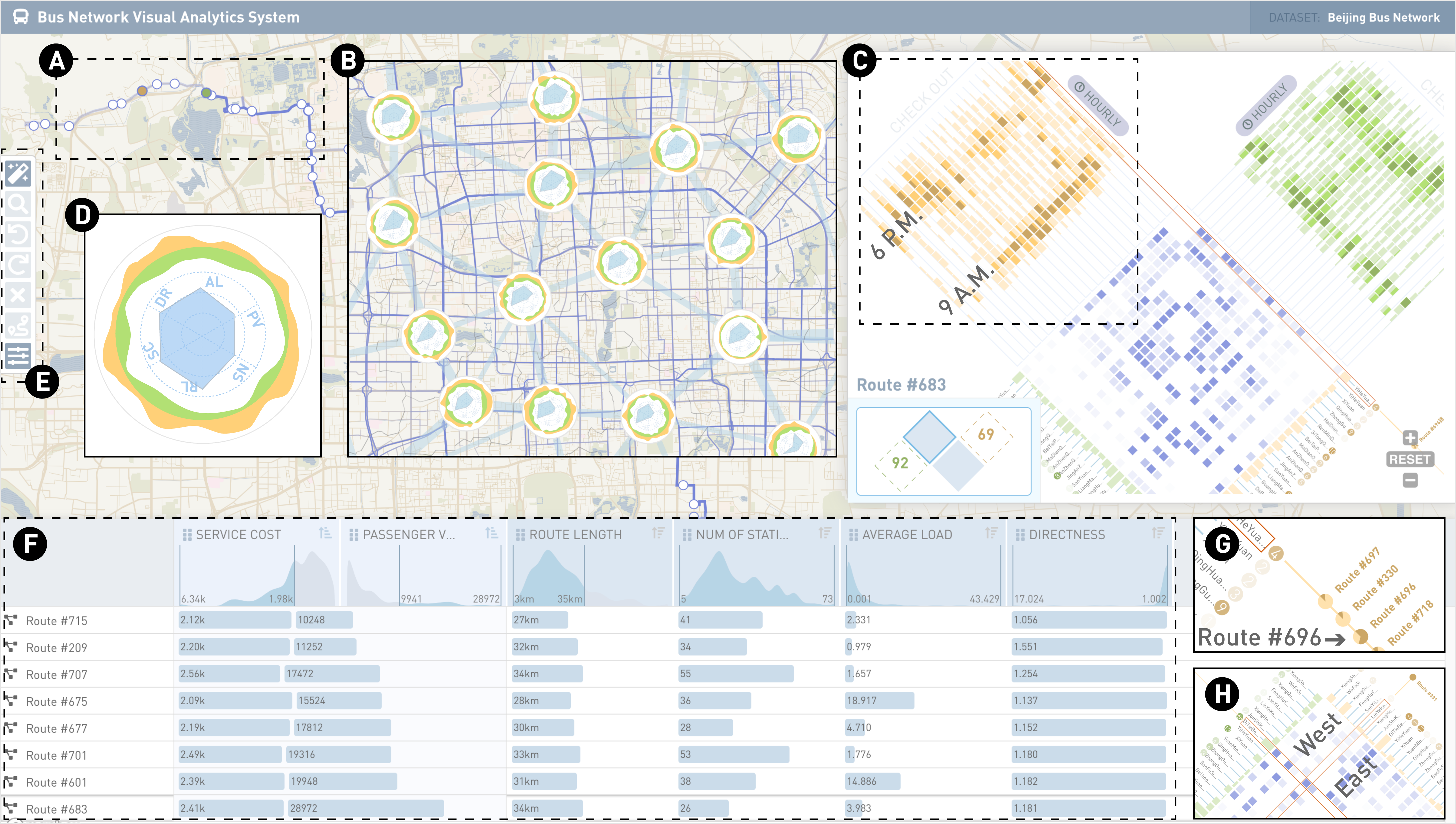}
  \caption{The interface of BNVA. A, C, G, H) The pattern indicated by the transfers from Route \#683 to \#696 suggests that Route \#683 can be extended farther into the west. B) The network-level analysis, comprising the map, route, and aggregation layers, facilitates the exploration of network performance. D) The zone glyph summarizes the statistics and passenger flows of a transportation zone. E) The toolbox provides fine-grained controls for the model. F) The route ranking view depicts the performance criteria of the routes.}
	\label{fig:teaser}
}

\vgtcinsertpkg

\ieeedoi{10.1109/TVCG.2020.3030458}

\begin{document}


\firstsection{Introduction}

\maketitle

In the last few centuries, the public transit bus service has become one of the most widely deployed public transportation backbones worldwide because of its flexibility and affordability.
Bus routes are typically updated every 3--5 years because the travel demand of bus riders is constantly changing~\cite{howplanned}.
However, revising bus routes remains difficult because a) it is hard to analyze the performance of a huge and complex bus network to determine which routes are problematic, and b) a large solution space constituted by many factors involved in the planning process~\cite{factors} needs to be extensively evaluated.
% for enabling planners to make judicious trade-offs.

Most bus networks are planned and updated manually or by analyzing small datasets based on planners' knowledge~\cite{BD549‐38,mees2010public,crisostomi_transportation_2018}.
Such methods can be laborious and time-consuming.
To identify feasible bus routes, data-driven planning methods~\cite{guihaire2008transit,mandl1980evaluation,pattnaik1998urban}, including the mathematical and heuristic ones, have been proposed to extract the routes based on predefined criteria.
However, most of these methods act as a black box, generating an optimized result based on the given input data and parameters without explanations.
Therefore, verifying the quality of the generated routes or adjusting the parameters to find an improved solution is difficult for domain experts.
The interpretability of route generation has been further studied by Chen et al.~\cite{DBLP:journals/tits/ChenZLZ14} and Weng et al.~\cite{8998192}.
% based on the extraction of Pareto-optimal transit routes.
Rather than producing a potentially unsatisfactory route, they propose to generate a set of \textit{Pareto-optimal transit routes}, where none of the routes is better than any other route for all given criteria.
% Although such methods can considerably reduce the size of the solution space,
However, experts are still required to manually compare among hundreds of routes and determine the most optimal one.

% \todo{Show a few studies on the automated methods for route planning.}
% \todo{What are the weaknesses of these automated methods?}

In this study, we collaborated with the experts from the transportation and urban computing domains to develop a solution that facilitates efficient analysis and incremental planning of bus routes.
Inspired by the urban data analytics studies~\cite{DBLP:journals/tvcg/LiuWL0ZQW17,DBLP:conf/chi/WengZ0ZW18}, we employ a visual analytics approach to integrate domain knowledge with the computational power of the Pareto-optimal route extraction method~\cite{8998192}, thereby enabling the experts to interactively interpret the performance of complex bus networks and evaluate the alternative routes generated in real-time.
Developing such a solution poses the following three challenges:

\vspace{1mm}
\noindent\textbf{In-depth analysis of complex bus route networks.}
The performance of bus route networks must be extensively evaluated first to detect the ineffective parts that need to be replanned.
However, bus route networks that operate in a large city may comprise a hierarchy of thousands of bus routes and stations, producing millions of trip records per month.
Analyzing such a complex dataset for deficient routes poses challenges in the scalability and effectiveness of the proposed solution.

\noindent\textbf{Interactive generation of improved route candidates.}
The Pareto-optimal route search model~\cite{8998192} can be used to generate potential route alternatives to the detected deficient routes.
However, the constraints provided by the model are insufficient for many practical scenarios, for example, planning the routes via a few key stops.
Moreover, since the model is progressive and time-consuming, the experts need to know when the results are sufficiently good to stop the search.
% These model issues poses the interactivity and usability challenges for the solution.
% This model was designed to be progressive, such that the model could produce better solutions if more time was given.
% However, the users cannot determine when the route candidates are good enough to stop the search without any indicators, posing the interactivity and usability challenges for the proposed solution.

\noindent\textbf{Effective evaluation of alternative bus routes.}
A large number of route alternatives will be generated with the model.
These alternatives can be diverse and complicated.
The experts request that they should be kept in the loop to make informed and transparent decisions.
However, the volume, diversity, and complexity of the generated candidates prohibit the experts from effectively evaluating these candidates based on topologies or performance criteria to identify the most optimal one.

% \todo{Describe how visual analytics can help address the bus route planning problem.}
% \todo{In this paper, we collaborated with domain experts and proposed a visual analytics solution that xxx based on large-scale bus trip data (the goal in one sentence).}
% \todo{Through the close collaboration, we followed the user-centric design procedure and managed to identified two major challenges in the problem of bus route planning:}

% \todo{\textbf{Where} to build bus routes?}
% \todo{\textbf{How} to build bus routes?}

\vspace{1mm}
To address these challenges, we propose BNVA (Bus Network Visual Analytics system), a novel visual analytics system that helps bus route planners analyze and improve the performance of bus route networks.
% , discovering outdated and abnormal bus routes, and manipulating route planning incrementally.
For the first challenge, we adopt a hierarchical exploration approach comprising network, route, and stop levels and propose a novel matrix view to facilitate the analysis of passenger flows and transfers;
for the second challenge, we extend the model to support various constraints and integrate tailored visualizations to depict the performance of the generated routes in real-time;
for the third challenge, we develop a progressive decision-making strategy based on route aggregation to help users evaluate the topologies and criteria of alternative routes.
The contributions of this study are as follows:

\begin{compactitem}
    \item We characterize the user requirements in analyzing and improving the performance of bus route networks;
    \item We extend the Pareto-optimal route generation model to support additional constraints from practical scenarios;
    \item We develop a novel visual analytics system for bus route planning, featuring a route matrix view for passenger flow analysis and a conflict resolution strategy for progressive route decision-making;
    % \item proposed a new route matrix view for hierarchical bus network exploration and a novel graph-based progressive planning procedure for alternative bus route assessment;
    \item We evaluate our approach with two usage scenarios based on a real-world dataset and received positive feedback from experts.
\end{compactitem}

% \todo{To address the aforementioned challenges, we developed BNVA, a novel visual analytics system xxx.}
% \todo{For the \textbf{where} challenge, we xxx.}
% \todo{For the \textbf{how} challenge, we xxx.}
% \todo{Our system was evaluated with the case studies based on the real-world data and has received positive feedback from the domain experts.}
% \todo{The contributions of this paper is summarized as follows:}

\section{Related Work}

This section summarizes the transit network design and urban data visual analytics studies that are closely related to the present study.

\subsection{Transit Network Design}

Transit network design is a huge and complex research topic in transportation research.
Guihaire et al.~\cite{guihaire2008transit} divided the process of transit network design and scheduling into three parts: transit network design problem (TNDP), transit network frequencies setting problem, and transit network timetabling problem.
This section mainly focuses on the related studies on the TNDP.

Many data-driven methods have been developed to generate transit networks automatically based on public travel demand, which can be determined from the survey~\cite{sun1998household,yu2005optimizing}, transportation~\cite{DBLP:journals/tits/ChenZLZ14}, and telecommunication~\cite{pinelli2016data} data.
These methods can be categorized into mathematical and heuristic methods.
Mathematical methods~\cite{mandl1979applied,murray2003coverage,guan2006simultaneous} generate the transit line configuration with numeric optimization approaches, including mixed-integer linear programming~\cite{cplex}.
Heuristic methods tackle the problem based on certain heuristics with the greedy approach~\cite{mandl1980evaluation}, heuristic search, and genetic algorithms (GAs), such as tabu search~\cite{zhao2008optimization}, GA variants~\cite{xiong1992transportation, chakroborty2002optimal}, simulated annealing~\cite{fan2006using}, and ant colony algorithm~\cite{yu2005optimizing}.
However, most of these methods only generate an optimized transit line configuration balancing multiple performance criteria with the given parameters.
To expand the decision-making scope with alternative solutions, Chen et al.~\cite{DBLP:journals/tits/ChenZLZ14} applied a random search approach in finding a set of transit routes that do not dominate each other in terms of two criteria, namely, route time and passenger flow, with the given origin and destination stops.
Weng et al.~\cite{8998192} elaborated on the definition of Pareto-optimal transit routes and proposed an efficient route extraction method based on the Monte-Carlo tree search.

Nevertheless, the aforementioned automated methods may generate unsatisfactory results and typically require users to laboriously analyze the results and adjust the input parameters accordingly.
Several preliminary efforts~\cite{reroute,optibus} in providing interactive solutions for the design procedure have been observed in the wild, but these solutions do not involve model-assisted visual planning of transit routes.
To the best of our knowledge, this study is the first step toward the in-depth visual analytics of interactive bus route analysis and adjustment.

\subsection{Visual Analytics of Urban Data}

Large-scale urban data, such as human mobility, social network, geographical, and environmental data, have been collected across cities in recent years with the advancement in data sensing and management technologies.
These urban data have made efficient data-driven solutions~\cite{DBLP:journals/tist/ZhengCWY14a} possible for various urban problems, such as bike lane planning~\cite{DBLP:conf/kdd/BaoHRLZ17}, ambulance location selection~\cite{DBLP:conf/gis/LiZJWUG15}, and travel time estimation~\cite{DBLP:conf/kdd/WangZX14}.
However, most of these methods cannot automatically produce satisfactory results~\cite{DBLP:journals/tvcg/LiuWL0ZQW17} without domain experts in the loop because of the complicated nature of urban problems.

To facilitate the analysis of urban data, many studies~\cite{DBLP:journals/tbd/ZhengWCQN16} have employed visual analytics that enables users to interactively obtain patterns and insights from complex datasets, typically with the aid of efficient computational models.
These studies mainly involve the visualization of time~\cite{DBLP:series/hci/AignerMST11} and location properties.
Time properties can be depicted with axis-based design~\cite{DBLP:journals/ivs/ZhaoFH08}, relative time encoding~\cite{DBLP:conf/ieeevast/WuZQCGN14}, and dynamic visual representations~\cite{gapminder}.
The visualization of location properties can be categorized into the point-, region-, and line-based techniques~\cite{DBLP:journals/tbd/ZhengWCQN16}.
Point-based techniques~\cite{DBLP:conf/ieeevast/AndrienkoAMMP10,passvizor,geo_point} represent the locations with the points in their spatial contexts.
Region-based techniques~\cite{DBLP:books/daglib/0032328,DBLP:journals/cgf/ZengFAQ13} render the aggregated location data based on certain spatial divisions.
Line-based techniques~\cite{DBLP:journals/tvcg/AndrienkoABDH13,DBLP:journals/vi/LiuJYTL19,DBLP:journals/cgf/ZengSJT19} encode the locations based on road or traffic networks with line-based representations.
Besides, other properties involved in the urban visual analytics studies, such as numbers and texts, can be presented with their corresponding visualization techniques~\cite{DBLP:books/daglib/0034520,DBLP:journals/tist/YangZQ16,DBLP:journals/vi/LiDY18}.
Furthermore, various techniques, including spatiotemporal~\cite{DBLP:conf/iv/GatalskyAA04,DBLP:conf/apvis/SunLWLQ14} and multivariate~\cite{DBLP:journals/tvcg/Keim02,DBLP:journals/tvcg/LiuMWBP17} visualizations, have been developed for the visualization of multiple properties combined.

With the aforementioned visualization techniques, urban visual analytics studies have explored various data sources, such as human mobility~\cite{DBLP:journals/tvcg/LiuWL0ZQW17,DBLP:conf/ieeevast/AndrienkoAHRW11,DBLP:journals/vi/NiSXQ17,DBLP:journals/tvcg/ZhouMTZGHC19}, social media~\cite{DBLP:journals/tvcg/CaoSLLLL16,DBLP:journals/tvcg/LiuLZLWP16}, environmental~\cite{DBLP:journals/tvcg/DengWCLWBZW20,DBLP:journals/tvcg/GoodwinDJDDDKSW13,9072315,DBLP:journals/cga/ZengY18}, and transportation data~\cite{DBLP:journals/tvcg/ZengFAEQ14,hughes2004visualization,DBLP:journals/cga/Pack10}.
Andrienko et al.~\cite{DBLP:journals/tits/AndrienkoACMZ17a} comprehensively summarized the prior studies on transportation data and categorized these studies into three directions, namely, transportation infrastructure~\cite{DBLP:journals/tvcg/AndrienkoAFW17, DBLP:conf/ieeevast/AndrienkoARNPG09}, transportation user behaviors~\cite{DBLP:journals/ivs/AndrienkoAFJ16,beecham2014exploring}, and transportation modeling and planning~\cite{DBLP:journals/ivs/AndrienkoAB08,DBLP:journals/tog/SewallWL11,customized_shuttle}.
Transportation data can be also used in studying place connectedness combined with novel visual analytics techniques~\cite{8758943}.
However, whether visual analytics can facilitate the efficient planning of bus transit routes remains to be elucidated.
In this study, we tackle this problem by combining the visualization techniques for the time and line-based location properties to provide an overview of bus network performance and enable users to interactively evaluate alternative bus routes.

\section{Background}

In this section, we show the formulation of the bus route planning problem and summarize the identified domain requirements.

\subsection{Problem Formulation}

In the past year, we collaborated with four domain experts (EA-ED) working on a bus network improvement project for a large city with over 7 million residents.
EA and EB are urban computing experts who have been studying sophisticated data-driven solutions for urban problems in the last few decades.
EC is a data scientist in charge of developing a route optimization model for the bus network improvement  project, and ED is the project manager of this project.
The goal is to find an improved bus route network that meets the travel demand of citizens.

An extensive literature review~\cite{guihaire2008transit} reveals that most data-driven methods focus on creating a new optimized route network from scratch.
However, such methods are impractical in our scenario because replanning the entire network will significantly disrupt commuters' daily routines.
By working with the domain experts, we constructed a three-stage iterative workflow for analyzing and improving bus networks:

% \begin{compactenum}
\vspace{1mm}
\noindent\textbf{Stage Exploration}:
At this stage, the experts aim to explore the performance of the bus network and determine ineffective routes.
The ineffective routes may exhibit several characteristics, such as \textit{low passenger flows} (few passengers will take these routes, leading to the waste of public transportation resources), \textit{high service cost} (these routes are costly to maintain mainly due to long service time and frequent scheduling), and \textit{poor route directness} (the transit distances on the buses are considerably longer than the road distances).
The analysis of the existing bus network is based on the historical bus trip records.

\noindent\textbf{Stage Manipulation}:
After an ineffective route is determined, the experts would like to obtain its replacement routes.
The Pareto-optimal route search method~\cite{8998192} is selected because of its capabilities in generating alternative routes on the approximated Pareto front based on multiple performance criteria.
However, the original method is progressive without indicating the quality of the current results in real-time.
Therefore, this stage aims to allow users to intuitively specify route generation constraints and determine when the generated routes are sufficiently good to terminate the search process.

\noindent\textbf{Stage Evaluation}:
Given a set of Pareto-optimal route alternatives, the experts aim to determine the improvement of the generated routes, make trade-offs between the performance criteria, and determine which alternative route is the most optimal.
Considering that each alternative route is a graph that shares some nodes with other routes, analyzing the similarities and differences between these routes with hundreds of the routes generated becomes increasingly challenging.
Hence, this stage aims to facilitate intuitive multi-criteria comparison and decision-making of alternative routes with tailored visualizations.
% \end{compactenum}

% \vspace{1mm}
% After the evaluation stage is completed, the users can either return to the exploration stage and start over to find more ineffective routes, or they can save the selected routes as a GeoJSON~\cite{rfc7946} file such that these routes can be exported into GIS softwares (e.g., ArcGIS~\cite{arcgis}) for further planning procedures like the location selection of bus stops and the scheduling of buses and drivers.

\subsection{Data Description}

The proposed system is based on three types of data, namely, bus stop, route, and trip data, collected from bus networks.

\begin{compactitem}
    \item \textbf{Bus stop data} comprise the bus stops in a city. Each stop is defined by its ID, name, and coordinates. %The stops are not placed at their accurate positions but at the center of the nearest road.
    \item \textbf{Bus route data} comprise the bus routes, where each of them is identified by its ID and a stop sequence.
    \item \textbf{Bus trip data} contain a series of bus fare card records, where each of them comprises a card ID, a tap-on timestamp, and the route and stops where the fare card was tapped on and off. However, the tap-off timestamps are not present in the dataset because of sensor errors. This timestamp was inferred either by using the tap-on timestamps at the destination stops if such transfer records exist, or based on driving time along the bus route at 20 km/h plus 2 min spent at every stop, as suggested by our domain experts.
\end{compactitem}

\subsection{Requirement Analysis}

To develop a feasible and practical approach for analyzing and improving bus networks with visual analytics, we followed the nested model for visualization design and validation~\cite{DBLP:journals/tvcg/Munzner09} to characterize the domain problem and identify the low-level analytical requirements.
Based on the three-stage workflow, we conducted literature reviews on the related studies, informal interviews with the experts, and brainstorming sessions with other visualization researchers to iteratively refine the workflow and compile the following analytical requirements.

\vspace{1mm}
\noindent\textbf{Stage Exploration (P).}

\noindent\textbf{P1: Obtain the spatial overview of the bus network and its performance.}
The experts requested to see a map-based overview of the bus network similar to other GIS software~\cite{optibus,arcgis}.
Such an overview should help users to rapidly orient themselves in the spatial context and grasp the spatial distribution of routes.
The overview should also include the visualization of the network performance to guide users in performing a drill-down analysis on the potentially ineffective routes.

\noindent\textbf{P2: Analyze the passenger flows of bus routes to find weaknesses.}
The movement of passengers through the bus network provides key insights for the experts to evaluate the performance of the routes.
For example, some parts of a route might be non-functional if few passengers were getting on or off in these parts.
Therefore, intuitive visualization of passenger flows is highly demanded to facilitate the identification of deficient routes.
Moreover, such a visualization should also enable the experts to analyze the transfers among multiple bus routes, which may reveal patterns to improve route design.

% The bus routes and their stops constitute a linear graph, and transfers among the routes form links between these graphs.
% The experts would like an interactive and intuitive visualization that facilitates the exploration of such nested graphs to understand the performance of these routes based on how passengers move through the bus network spatiotemporally.
% The design should also enable users to efficiently identify the routes that need to be improved and designate these routes as the input of the manipulation stage.

\vspace{1mm}
\noindent\textbf{Stage Manipulation (M).}

\noindent\textbf{M1: Generate a set of alternative routes based on the constraints.}
The experts prefer to interact with the model and generate the Pareto-optimal routes as the potential replacements of the selected ineffective route.
To minimize the disruption caused by route changes, the experts may preserve several primary stops from the original route.
Moreover, certain constraints, such as the construction cost and maximum route length, may be imposed on route generation.
An easy-to-use interface is required for translating these constraints into the complex parameters required by the model.

\noindent\textbf{M2: Inspect the quality of the generated routes in real time.}
Considering that the model is progressive and does not stop automatically, the experts need to know when the results are sufficiently good to stop the route generation process.
Hence, tailored visualizations are required to depict the current status of the generation process in real-time and provide the early quality preview of the alternative routes as the process continues.
Moreover, such visualizations should allow some undesired routes to be removed from the search space to interactively guide and accelerate the search process.

\vspace{1mm}
\noindent\textbf{Stage Evaluation (L).}

\noindent\textbf{L1: Compare the generated routes based on topologies.}
To help the experts identify the most promising ones from the generated routes, the proposed system must reveal the topological similarities and differences among these routes.
The experts may want to know: \textit{What stops do these two routes share? Which pair of consecutive stops is the most frequently selected? How much does a route deviate from another one by taking a detour?}
Integrating topological information can help experts estimate the performance of these routes and eliminate the undesired ones that share similar characteristics.

\noindent\textbf{L2: Compare the generated routes based on multiple criteria.}
The most promising alternative route should also be determined in terms of the performance criteria.
However, experts may treat each criterion differently under different circumstances.
For example, the distances between the stops in a suburban bus route will be considerably longer than those between the stops in a city bus route.
To facilitate a judicious decision-making process, the system should enable the experts to inspect the criteria of the generated routes and identify the most optimal one efficiently with tailored ranking models.

\section{Model}

\label{sec:model}
In this section, we introduce the route generation method that supports the visual analytics of bus route planning and propose the modification and improvements to tailor this method for the interactive analysis.

\subsection{Pareto-Optimal Route Search}

This subsection summarizes the core idea of Weng et al.'s method~\cite{8998192} that searches Pareto-optimal transit routes based on the Monte-Carlo search tree.
The Monte-Carlo search tree~\cite{DBLP:journals/tciaig/BrownePWLCRTPSC12} is studied to search the best next move in a game.
Starting from a given game state, the search repeats four stages, namely, selection, expansion, simulation, and backpropagation.
First, the most promising state is selected.
Then, a new state is created based on the estimated best next move of the selected state.
Next, the game result is obtained via simulation.
Finally, the estimated value of the states in the tree is updated accordingly.

Given an origin and a destination station $s_o$ and $s_d$ in a set of stations $S=\{s_1, s_2,...\}$, a feasible transit route between the origin and destination stations is denoted as $R=\{r_1,r_2,...,r_n\}\in P_{od}$, where $r_1=o$, $r_n=d$, $s_{r_i}\in S$, and $P_{od}$ comprises all routes between $s_o$ and $s_d$.
A set of criteria $C=\{c_1,c_2,...\}$ can be defined to measure the performance of a route $R$.
% If $R_i$ is \textit{strictly better} than $R_j$ with respect to. the criterion $c_k$, we say $c_k(R_i)>_{c_k}c_k(R_j)$.
We say $R_i$ \textit{dominates} $R_j$ if a) $R_i$ is better than or equal to $R_j$ with respect to every criterion $c\in C$ and b) $R_i$ is strictly better than $R_j$ with respect to at least one criterion.
The goal is to find an optimal Pareto-set of feasible transit routes $P=\{R_1,R_2,...\}\subseteq P_{od}$, where $R_i$ is \textit{not} dominated by $R_j$ for any $R_i, R_j\in P$.

\begin{figure}[!tb]
    \centering
    \includegraphics[width=0.48\textwidth]{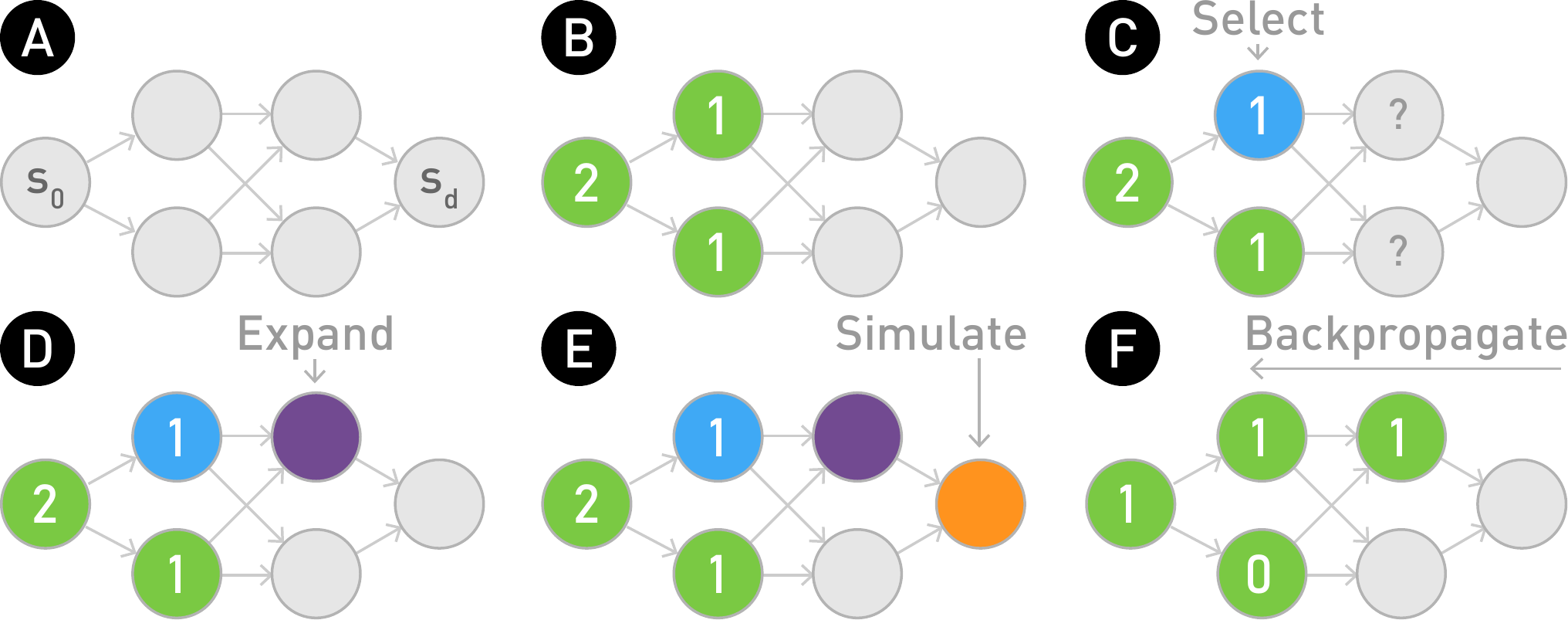}
    \caption{A) A station graph generated between stations $s_0$ and $s_d$. B) An example of the state tree (green nodes represent the explored states, and the numbers of the discovered Pareto-optimal routes are shown in the nodes). C) The most promising state (blue) is selected at the selection stage. D) The most promising neighboring station (purple) is expanded at the expansion stage. E) The simulation begins from the expanded station to $s_d$. The involved stations are marked in purple. F) The route obtained from the simulation is tested against the current set of Pareto-optimal routes, and the numbers in the nodes are updated.}
    \label{fig:model}
\end{figure}

Initially, a directed acyclic station graph $G_{od}=(S,E_{od})$ (Fig.~\ref{fig:model}A) is constructed between the origin and destination stations $s_o$ and $s_d$ with the approach proposed by Chen et al.~\cite{DBLP:journals/tits/ChenZLZ14}, where the edges $E_{od}$ among the station nodes $S$ are given based on heuristic criteria.
A search strategy is then proposed based on the following five stages.
% , which are similar to those given in the Monte-Carlo tree search method~\cite{DBLP:journals/tciaig/BrownePWLCRTPSC12}:

\textit{Initialization.}
A state tree is initialized with a root node representing the origin station.
A set of discovered Pareto-optimal transit routes $P_s$ is maintained throughout the search process.
Fig.~\ref{fig:model}B shows an example state tree.
The numbers in the nodes indicate the numbers of the discovered Pareto-optimal transit routes involving the nodes.
%  and associates each node $m$ in the state tree with the number of times that the corresponding station has appeared in $P_s$, denoted by $\zeta_m$, and the number of times that the node has been visited, denoted by $\omega_m$.

\textit{Selection.}
This stage aims to find the most promising node $m_p$.
Starting from the root node, the search process recursively descends in the state tree by selecting a child node with the maximum upper confidence bound~\cite{DBLP:conf/ecml/KocsisS06}, balancing between the exploration of the less-visited and high-valued nodes.
The selection stops if any neighboring station of the current node has not been added to the state tree.
The selected node $m_p$ is marked in blue in Fig.~\ref{fig:model}C.

% determines whether the current node has been \textit{fully-expanded}, i.e., all the neighboring stations have been added to the state tree.
% If the node is fully-expanded, one of its child nodes with the maximum $\zeta$-UCB value is selected. $\zeta$-UCB is adapted from the upper confidence bound (UCB) formula~\cite{DBLP:conf/ecml/KocsisS06} and can be computed as follows:
% \begin{equation*}
%     \zeta\text{-UCB}(m_q)=\frac{\zeta_{m_q}}{\omega_{m_q}}+\lambda\sqrt{\frac{\ln\omega_{m_p}}{\omega_{m_q}}},
% \end{equation*}
% where $m_p$ is the parent of $m_q$ and $\lambda$ is a parameter that controls how much the less-explored nodes are preferred to the high-valued ones.
% The selection terminates if the current node is not fully-expanded.

\textit{Expansion.}
This stage seeks to identify the most promising neighboring station of $m_p$ that is not in the state tree.
All routes belonging to a node in the state tree begin with a \textit{route prefix} $R^k=\{r_1,...,r_k\}$ and constitute a \textit{route subspace} $P_{od}(R^k)$.
The search process estimates the gain $\Delta c_i$ with respect to every criterion $c_i\in C$ in choosing a neighboring station $s_{e_i}$ and its associated route subspace $P_{od}(R^k\cup \{e_i\})$ based on heuristic approaches.
The details of these approaches can be found in the original paper.
Thereafter, the gains are normalized and averaged for each neighboring station, and a neighboring station is randomly chosen based on the distribution of the gains.
This station, marked in purple in Fig.~\ref{fig:model}D, is inserted into the state tree as a child node of $m_p$.

% The total average gain $G(e_i)$ can be written as follows:
% \begin{equation*}
%     G(e_i)=\frac{1}{|C|}\sum_{c_i\in C}\mathcal{N}_{c_i}(\Delta_{c_i}(P_{od}(R^k_u\cup \{e_i\}))),
% \end{equation*}
% where $\mathcal{N}_{c_i}$ normalizes the average gain with respect to. the criterion $c_i$ to $[\epsilon, 1]$ ($\epsilon$ is a predefined minimum normalized gain).
% The probability $\text{Prob}(e')$ of choosing a neighboring station $s_{e'}$ is given as follows:
% \begin{equation*}
%     \text{Prob}(e')=\frac{G(e')^\alpha}{\sum_{e_i}G(e_i)^\alpha},
% \end{equation*}
% where $\alpha$ controls how much the search process is biased towards the stations with higher gains.
% Hence, a neighboring station can be randomly chosen based on the probability distribution given by this formula.
% The chosen station is then inserted into the state tree as a child node of $m_p$.

\textit{Simulation.}
Starting from the newly inserted node, the search process recursively performs a similar selection strategy to that in the expansion stage until the destination station is reached (the simulation nodes are marked in orange in Fig.~\ref{fig:model}E).
Then, the obtained route is tested against the Pareto-optimal transit route set $P_s$.

\textit{Backpropagation (Fig.~\ref{fig:model}F).}
If the obtained route is Pareto-optimal, $P_s$ will be updated to remove the routes that are no longer Pareto-optimal.
The search process returns to the selection stage or terminates when the time limit is exceeded or the station graph is exhausted.

% Given a set of viable stations $S=\{s_1, s_2, ...\}$, an origin and a destination station $s_o$ and $s_d$, the goal of this method is to find a set of routes $P=\{R_1,R_2,...\}$

% The method comprise a two-fold strategy.
% The first step is to build an accurate prediction model that estimates the value of each

\subsection{Model Optimization}

This subsection introduces the optimization and improvements we have implemented for the Pareto-optimal route search model to make it suitable for the interactive analysis.

\subsubsection{Efficient Computation of Route Directness}

Route directness~\cite{zhao2004transit}, which measures how much the transit network distance between each pair of stations differs from the corresponding shortest road network distance, is a crucial factor to consider in the bus route planning process.
High route directness typically indicates that buses take minimal detours, and passengers will spend minimal time in traveling on this route.
% In the expansion stage, Weng et al.~\cite{8998192} proposed a simple yet effective averaging heuristic that estimates the route directness $I_R$ of a route subspace $P_{od}(R^k)$ constituted by all routes starting with prefix $R^k=\{r_0, r_1, ..., r_k\}$ as follows:
% \begin{equation*}
%     I_R(P_{od}(R^k))=\frac{1}{N_{r_kd}}\sum_{v=\eta(s_{r_k})}^{\eta(s_d)}((\sum_{u\in R^k}\frac{\delta_{uv}}{D_{uv}})+(\sum_{u=\eta(s_{r_k})+1}^{v-1}\frac{\delta_{uv}}{D_{uv}})),
% \end{equation*}
% where $\eta(s_i)$ denotes the topologically sorted index of station $s_i$ in the station graph, $N_{uv}$ is the number of possible paths between stations $s_u$ and $s_v$, and $\delta_{uv}$ and $D_{uv}$ are the transit network and shortest road distances, respectively, between stations $s_u$ and $s_v$.
In the expansion stage, Weng et al.~\cite{8998192} proposed an averaging heuristic that estimates the route directness $I_R$ of a route subspace $P_{od}(R^k)$ constituted by all routes starting with prefix $R^k=\{r_0, r_1, ..., r_k\}$.
However, computing such a heuristic can be time-consuming because the time complexities are estimated to be $O(|S|^2)$ for the expansion stage and $O(|S|^3)$ for the simulation stage.
To accelerate the computation, we compute $I_R$ with the following formula:
\begin{eqnarray*}
    I_R(P_{od}(R^k))=\frac{1}{N_{r_kd}}((\sum_{u\in R^k}A(u, r_k))+B(r_k)), \\
\end{eqnarray*}
where $A(p,q)=\sum_{v=\eta(s_{q})}^{\eta(s_d)}\frac{\delta_{pv}}{D_{pv}}$, and $B(q)=\sum_{v=\eta(s_{q})}^{\eta(s_d)}\sum_{u=\eta(s_{q})+1}^{v-1}\frac{\delta_{uv}}{D_{uv}}$.
$\eta(s_i)$ denotes the topologically sorted index of station $s_i$ in the station graph, $N_{uv}$ is the number of possible paths between stations $s_u$ and $s_v$, and $\delta_{uv}$ and $D_{uv}$ are the transit network and shortest road distances, respectively, between stations $s_u$ and $s_v$.
$A(p,q)$ and $B(q)$ can be efficiently precomputed for every $s_p,s_q\in S$ because they are unrelated to route prefix $R^k$, resulting in low time complexities $O(|S|)$ and $O(|S|^2)$ at the expansion and simulation stages, respectively.

\begin{figure*}[!t]
    \centering
    \includegraphics[width=0.92\textwidth]{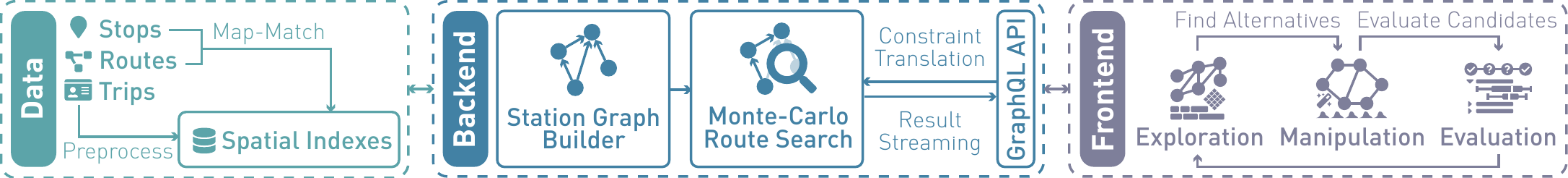}
    \caption{The system architecture of BNVA. BNVA comprises three parts, including data storage, backend, and frontend. Data storage preprocesses the data and indexes the data with a spatial database. Backend builds station graphs and answers route generation requests. Frontend combines the exploration, manipulation, and evaluation interfaces to establish an iterative workflow for analyzing and improving the performance of bus networks.}
    \label{fig:architecture}
\end{figure*}

\subsubsection{Translation of Generation Constraints}

To effectively integrate the route generation model with the proposed system, the experts requested us to implement practical route constraints, including multi-stop anchoring, construction cost, and service cost, in addition to route service time, demand satisfaction, and route directness defined in the original model.

\textit{Multi-stop anchoring.}
The model only allows users to define the origin and destination stops.
However, experts prefer to anchor several primary stops of the original route to minimize the impact on route network structure and riders' daily routines caused by route changes.
Moreover, such functionality is particularly useful when only a part of a route needs to be adjusted.
To achieve multi-stop anchoring, we reformulate the input of the model as a set of stop sets $\mathbb{S}=\{S_1=\{s_{11},s_{12},...\}, S_2, ...\}$ and construct a station graph by combining the subgraphs generated in the following procedure.
For each stop set $S_i\in\mathbb{S}$, the stops in this set will be sequentially connected as a station subgraph; and for each pair of consecutive stop sets $S_i,S_{i+1}\in\mathbb{S}$, a station subgraph will be generated between the last stop in $S_i$ and the first stop in $S_{i+1}$.
After the full station graph is built, we test each node in this graph against the criteria proposed by Chen et al.~\cite{DBLP:journals/tits/ChenZLZ14} and remove the nodes that neither belong to any stop set in $\mathbb{S}$ nor satisfy the criteria.
Therefore, the Pareto-optimal routes generated on this graph contain all the stops in their order given in $\mathbb{S}$.

\textit{Construction cost.}
Integrating the construction cost enables experts to specify the cost of establishing a bus route as the estimated cost per stop $C_s$ times the number of stops in this route.
We propose to add a new construction cost criterion $C_R$ that measures the average construction cost of a route subspace $P_{od}(R^k)$ with route prefix $R^k=\{r_1,...,r_k\}$.
The number of possible routes from a stop $s_i$ to destination stop $s_d$, denoted by $n_i$ can be computed by traversing in the station graph from $s_i$ to $s_d$.
Iterating from $s_d$ back to $s_{r_k}$, we have: $c_i=\frac{1}{n_i}\sum_{j\in E_i}n_{j}(c_{j} + C_s)$, where $E_i$ comprises the indices of the neighboring stops of $s_i$.
Hence, we have $C_R(P_{od}(R^k))=c_{r_k}$.
The computation of this formula is efficient because the time complexity is estimated to be $O(|S|)$.

\textit{Service cost.}
Service cost indicates the cost of scheduling and operating a bus route.
The estimation of service cost is based on several factors including headway $H$ (the time interval between two consecutive buses), service span $T_s$ (the period when the route is operating), one-way route service time $T_R$, crew wage per hour $C_w$, fuel cost per kilometer $C_f$, and maintenance cost per kilometer $C_b$.
We choose $v=20\text{km/h}$ to be the average speed of buses, as recommended by the experts.
Service cost $C_v$ can be derived and estimated from the route service time criterion $T_R$: $C_v=\frac{T_s}{H}(2T_RC_w+2T_Rv(C_b+C_f))$.

\subsubsection{Parallelized Pareto-Optimal Route Search}
We accelerate the route generation model by exploiting multiple cores on modern workstations and parallelizing the search process.

In the expansion stage, rather than choosing a neighboring station randomly based on gain distributions, we sort the neighboring stations by their average gains and select the largest-$k$ stations, where $k$ is the maximum number of threads to use specified by the user.
Multiple simulation stages, which still use the original random strategy, are spawned in parallel for these stations.
The generated candidate routes are then collected and tested against the Pareto-optimal transit route set in batch at the backpropagation stage.
%To reduce the thread spawning overhead for each search iteration, we maintain a thread pool and allocate workers for the simulation stage from the pool.

\subsubsection{Model Interactivity}

We propose the following modifications to improve the interactivity and responsiveness of the aforementioned model and provide better integration with the visual analytics interface.
% To better integrate the aforementioned model with the visual analytics interface, we propose the following modifications to improve the interactivity and responsiveness of the model.

% \textit{Result streaming.}
% To visualize the route generation process in the system and support real time visual analytics, we expose the internal states of the Monte-Carlo search tree and the route adding and removing changes in the Pareto-optimal transit route set via a WebSocket~\cite{rfc6455} endpoint.
% The clients can subscribe to this endpoint and retrieve the streaming updates on the generation process.

\textit{Interactive search.}
Based on the current route generation result, the users may find some stations unnecessarily included in the routes and prefer to manipulate the search space by removing such stations from the station graph.
Moreover, the model should support adding stations that are not included in the station graph.
The model processes such requests by first regenerating the station graph incrementally and then altering the state tree and the route set to reflect the changes.

\textit{Route pruning.}
The route generation model may produce a sheer volume of Pareto-optimal routes because of the increasing number of performance criteria.
However, many routes can be unsatisfactory because they may perform well in some criteria but extremely poor in other ones.
To eliminate these routes, we allow users to specify the desired value range for each criterion.
As the search process computes the estimation heuristics (e.g., route directness $I_R(P_{od}(R^k))$), we use a similar heuristic strategy to estimate the minimum and maximum criterion values for each neighboring station.
A station will not be chosen if its estimated value range does not overlap with the specified one.
Such a pruning strategy not only integrates criterion filters into the model but also improves the efficiency of route generation.

% \begin{compactitem}
%     \item Route directness efficiency
%     \item ... (parallelization?)
% \end{compactitem}

% \subsection{Additional Constraints}

% \begin{compactitem}
%     \item Multiple stops
%     \item Number of stops
%     \item Construction budget
% \end{compactitem}

% \subsection{Interactivity}

% \subsubsection{Result Streaming}
% \subsubsection{Interactive Search}
% \subsubsection{Route Filtering}

\section{System Architecture}

BNVA is a web-based visual analytics application constituted by three parts, namely, data storage, backend, and frontend (Fig.~\ref{fig:architecture}).
% The architecture of BNVA is illustrated in Fig.~\ref{fig:architecture}.
The data storage preprocesses data sets and indexes them spatially in the database.
%us stop, route, and trip data, map-matches the data with Open Source Routing Machine~\cite{luxen-vetter-2011}, and indexes them spatially with a PostGIS database.
The backend handles route generation requests and exposes the internal states and APIs of the generation model to the frontend.
The frontend comprises three visual interfaces, namely, the exploration, manipulation, and evaluation interfaces.
The exploration interface facilitates the performance analysis of the existing bus network, the manipulation interface enables users to interact with the progressive model, and the evaluation interface assists users in comparing among the candidate routes based on topologies and performance criteria.
% to determine the best one.

\section{Visual Design}

To facilitate efficient exploration and improvement of bus route networks, BNVA comprises three visual interfaces, namely, the exploration (P1, P2), manipulation (M1, M2), and evaluation (L1, L2) interfaces.
The visual information-seeking mantra~\cite{DBLP:conf/vl/Shneiderman96} motivates us to develop an exploration interface that empowers the users to discover deficient routes in the network from the overview with a map (P1) to the details with novel flow matrices depicting passenger flows and transfers (P2).
The alternative routes can be obtained via the manipulation interface (M1), which provides tailored visualizations of route performance to informed the users of the generation progress and the quality of route replacements in real-time (M2).
The evaluation interface adopts a \textit{conflict resolution} strategy that iteratively extracts the topological differences and route clusters from the generated routes to facilitate the progressive and reliable decision-making process in finding optimal routes based on topologies (L1) and performance criteria (L2).

% Thereafter, the manipulation interface (M1, M2) is invoked for generating alternatives for the selected defective route.
% The progressive visualization provided by this interface enables the users to reliably obtain and evaluate streaming results from the route generation model.
% To find the most promising alternative, the evaluation interface (L1, L2) extracts the \textit{conflicts} between the generated routes, which can be resolved interactively via a set of tailored visualizations, thereby allowing the users to efficiently characterize hundreds of alternatives and determine an optimal replacement route judiciously based on their domain knowledge.

% , with spatial aggregation of bus stops, multi-criteria ranking of the routes, and a novel route matrix view which facilitates the in-depth analysis of passenger flows and route relationships (P1, P2).
% After a route is determined for optimization, the manipulation interface (Fig.~xx) streams the alternative routes from the route generation model in real-time and visualizes the routes on the map and in the ranking view for progressive analysis (M1, M2).
% Adopting a \textit{conflict-merge} strategy, the evaluation interface (Fig.~xx) helps the users compare among the route subspaces, generated from the alternative routes, with the map, ranking, and matrix views and build the optimal route interactively by choosing the most promising route spaces (L1, L2).

\subsection{Exploration Interface}
A large-scale bus network typically comprises a complex hierarchy constituted by massive routes, stops, and trips.
%The sheer volume of bus trips in the network also considerably contributes to the complexity of such a hierarchy.
To facilitate the identification of deficient routes, we organize the interface by the network-, route-, and stop-level analyses.
For the network-level analysis, a spatial aggregation view is designed to provide a spatial overview of the entire network and help the users filter routes with spatial constraints (P1, P2).
For the route-level analysis, a route ranking view is implemented to depict the performance of the routes (P1), allowing the users to find inefficient routes based on performance criteria.
For the stop-level analysis, a route matrix view is proposed to visualize the passenger flows and transfers among the stops in a selected route with matrices, establishing fine-grained inspection of route performance (P2).
% Three levels of analyses combined will support the informed decision-making in the identification of deficient routes.

% perform the graph-based analysis on route performance (P2).

% The exploration interface (Fig.~xx) comprises a spatial aggregation view, a route ranking view, and a route matrix view.
% The spatial aggregation view groups bus stops spatially on the map for an overview of the entire network (P1), facilitating intuitive filtering and selection of the routes.
% The route ranking view depicts the performance of the routes with the table-based ranking visualization~\cite{DBLP:journals/tvcg/GratzlLGPS13,DBLP:journals/tvcg/WengCDWCW19}, allowing the users to find interesting routes by ranking them.
% The route matrix view visualizes the check-in/out records and passenger flows of a selected route with matrices, enabling the users to perform the graph-based analysis on the performance of the selected route at a fine granularity (P2).

\subsubsection{Network-Level Analysis}
An overview of the bus network is essential in locating the areas where the inefficient routes most likely exist.
The spatial aggregation view (Fig.~\ref{fig:teaser}B), designed for the network-level analysis, comprises three linked layers, namely, the map, route, and aggregation layers, to help users analyze the network on a broad scale.

The map layer comprises a base map.
% with the Mapbox GL library~\cite{mapboxgl}.
The route layer draws all routes on the map in blue with opacity.
% To show the spatial distribution of the routes, the route layer draws all routes in the bus network on the map in blue with opacity, thereby encoding the number of routes with the density of the color.
However, the route layer cannot depict the network topology because of the overlapping routes.
Therefore, we designed the aggregation layer to visualize the topology with an \textit{aggregation graph}.
Each node in the aggregation graph corresponds to a group of bus stops aggregated spatially with hierarchical clustering~\cite{johnson1967hierarchical} by balancing the number of stops in each group.
%, which can be adjusted in accordance with the users' preferences.
These groups divide the city into \textit{transportation zones}.
The boundaries of these zones are computed with a Voronoi diagram~\cite{fortune1987sweepline} by unifying the polygons that enclose the bus stops inside the zones.
In addition, the numbers of the routes between the zones are encoded with the link widths.

A \textit{zone glyph} (Fig.~\ref{fig:teaser}D) is placed at the centroid of each transportation zone to summarize the key statistics of this zone.
The radar chart at the glyph center encodes six averaged criteria, namely, route length (RL), number of stops (NS), passenger volume (PV), average load (AL), route directness (DR), and service cost (SC).
Users can configure the visibility and order of the dimensions flexibly in the context menu.
% These averages are computed based on the parts of the routes covered by the zone.
% Such a chart can quickly give the users an overview of the performance of the corresponding part in the bus network.
Two diverging circular distributions around the glyph visualize the amount of passenger flows by the geographical directions in which the passengers in this zone leave (green) or enter (orange).
Double clicks magnify the glyphs, allowing users to obtain a clearer view of the radar charts inside.
The design of this glyph is kept simple yet informative, such that the users can naturally obtain and compare the performance of different zones with a number of glyphs.
%By hovering and clicking on the glyphs, the users can also filter the routes and keep those passing through the selected zones in the view.
%, the boundaries of which are visualized based on the voronoi diagram computed with the involved bus stops.

\subsubsection{Route-Level Analysis}
In the route-level analysis, we focus on assisting the users in finding inefficient routes based on performance criteria in complementary to the spatial information.
Inspired by LineUp~\cite{DBLP:journals/tvcg/GratzlLGPS13}, a table-based ranking visualization is included in the route ranking view to facilitate the multi-criteria analysis of the routes.
As illustrated in Fig.~\ref{fig:teaser}F, the columns represent six criteria similar to the zone glyphs, and each row is a route that can be ranked by selecting any column.
The criteria can also be grouped and sorted with different weights.
%, enabling the users to effectively analyze the weighted sum of these criteria and construct tailored ranking models.
In addition, the criterion distributions are shown in the column headers, providing an overview and range filters for the routes in the table.

% adjusting the widths of the columns and sorting the routes by several columns combined enable the users to analyze the weight sum of multiple criteria~\cite{triantaphyllou2000multi}.
% In addition, a value distribution is placed inside each column header to help the users effectively grasp a performance overview of the routes in the table and rule out unwanted routes by specifying value ranges.
% In each column header, a distribution of the corresponding criterion

\subsubsection{Stop-Level Analysis}
The stop-level analysis enables the users to explore and evaluate the passenger flows and transfers among the stops in a selected route.

\begin{figure}[!tb]
    \centering
    \includegraphics[width=0.48\textwidth]{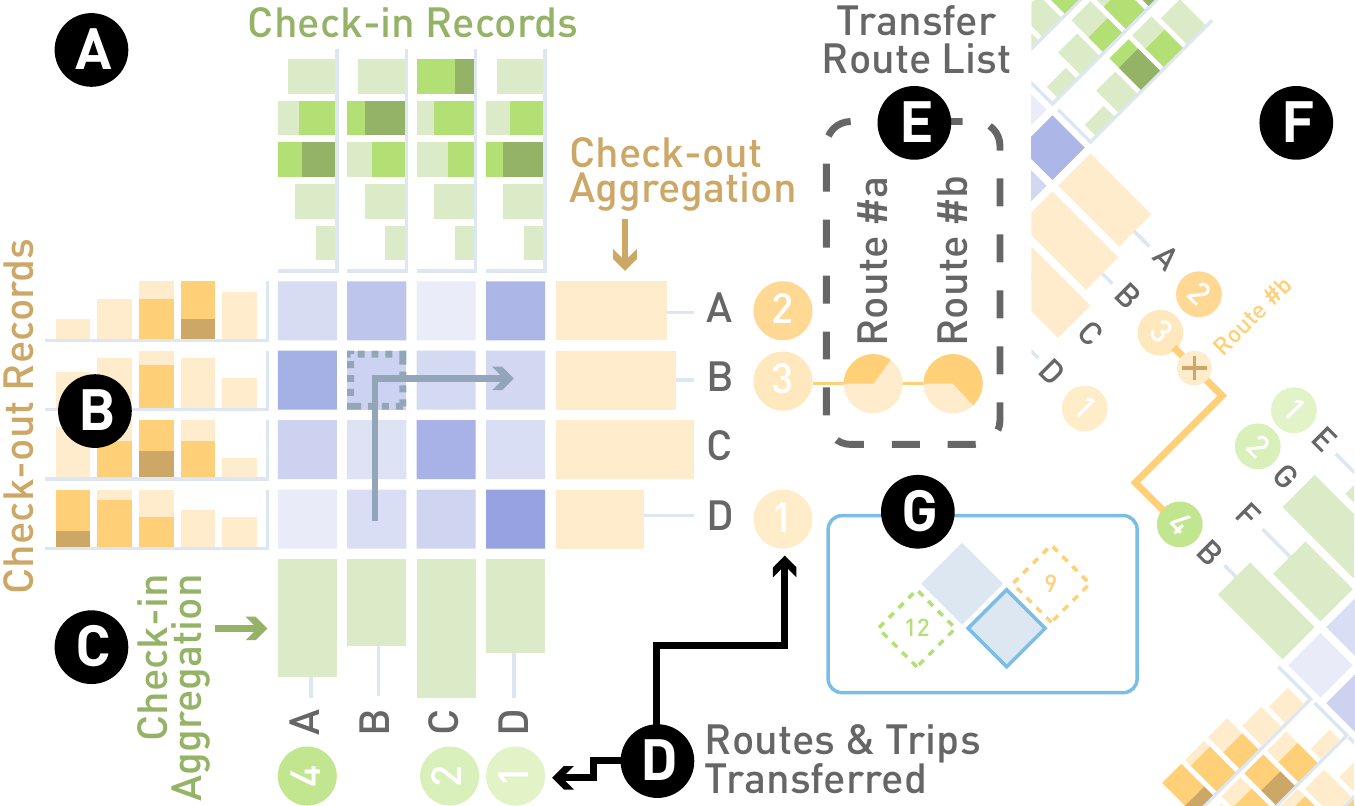}
    \caption{The design of flow matrices. A) Flow matrices visualize the flows and transfers of passengers on the selected routes. B) The horizon charts show the temporal distribution of check-in/out records. C) The bar charts aggregate the total passenger flows. D) The numbers of transfer trips and routes are encoded with the opacity of the circles and the numbers on them. E) Clicking on the circles brings up a list of transfer routes. F) A new flow matrix for the transfer route will be aligned and linked with the original one. G) The overview of all flow matrices.}
    \label{fig:matrix}
\end{figure}

A \textit{flow matrix} (Fig.~\ref{fig:matrix}A) is designed to visualize the passenger flows and transfers.
The columns and rows of the matrix correspond to the bus stops, and the color intensity of each cell in the matrix encodes the number of passengers traveling between the stops.
%from the column stop to the row stop.
The color intensity is computed by normalizing the number of passengers into $[0, 1]$ against a global passenger flow threshold, which can be changed by users.
%  represented by the column (the \textit{check-in stop} where the passenger enters the bus) to the row stop represented by the row (the \textit{check-out stop} where the passenger leaves the bus).
The horizon charts (Fig.~\ref{fig:matrix}B) visualize the number of passengers, aggregated by time, checking in or out at the corresponding stops.
The visibility of the horizon charts can be toggled in the context menu to simplify the view.
Three-band horizon charts are chosen over bar charts because the estimation accuracy of horizon charts is better than that of bar charts when chart height is limited~\cite{DBLP:conf/chi/HeerKA09}.
In addition, the bars (Fig.~\ref{fig:matrix}C) encoding the number of passengers per stop are positioned to the bottom and right of the matrix.
In case of long bus routes, users can right click on the matrix to select stops they wish to keep in the view.

Massive transfers may indicate that the routes are not well planned and discourage the use of bus transportation~\cite{zhao2006large}.
To visualize transfer information, we encode the number of passengers transferred to or from other routes with the opacity of the circles (Fig.~\ref{fig:matrix}D) next to the station names.
The numbers on the circles indicate how many routes passengers have transferred to or from.
The minimum opacity of the circles has been tuned to maintain the visibility of the numbers inside them.
% The numbered circle at the end of each row or column indicates the number of the routes passengers have transferred to or from at the corresponding stop.
Clicking on a circle expands a list of the associated routes (Fig.~\ref{fig:matrix}E), each preceded by a small pie chart indicating the percentage of passenger flows transferred to or from the route.
Selecting a route in the list reveals another flow matrix visualizing this route, which is aligned and linked to the original matrix (Fig.~\ref{fig:matrix}F).
% The new matrix, which is aligned and linked to the original one, can be toggled between showing the full or only transferred passenger flows.
Inspired by MatrixWave~\cite{DBLP:conf/chi/ZhaoLDHW15}, we rotate the matrices by 45\si{\degree} clockwise to accommodate them linearly for enhanced scalability.
An overview of the matrices (Fig.~\ref{fig:matrix}G) is provided at the bottom-left of the view, where each matrix is represented with a square.
The currently focused matrix is outlined in the overview, and the numbers of transferred routes are enclosed in the dashed squares.
% In addition, the average load and passenger flows per stop of the currently-focused route are visualized with the line chart to the right of the overview and the node-link diagram on the map, repsectively.

% Inspired by MatrixWave~\cite{DBLP:conf/chi/ZhaoLDHW15}, the matrix is rotated 45 degree clockwise to

% After the users select a route in the ranking view, the route matrix view, which facilitates the graph-based analysis of passenger flows for the select route, will appear at the right of the system interface.
% The passenger flows are depicted with a matrix rotated 45 degree clockwise.

% The route matrix view is designed to facilitates the graph-based analysis of passenger flows for the route chosen in the ranking view.
% After the users select a route in the ranking view, the matrix view will appear at the right of the system interface, presenting a flow matrix rotated 45 degree clockwise.
% Each row or column of the matrix represents a bus stop, and each cell encodes the amount of the passenger flows from the column stop (i.e., the check-in stop) to the row stop (i.e., the check-out stop).
% Each row or column also corresponds to a horizon chart placing to the left or top of the matrix, visualizing the frequencies of the check-out or check-in records occurred at the corresponding stop.
% Using the horizon charts instead of bar charts
% After a route is selected in the route ranking view, the route matrix view will appear and provide

\subsection{Manipulation Interface}

\begin{figure}[!tb]
    \centering
    \includegraphics[width=0.48\textwidth]{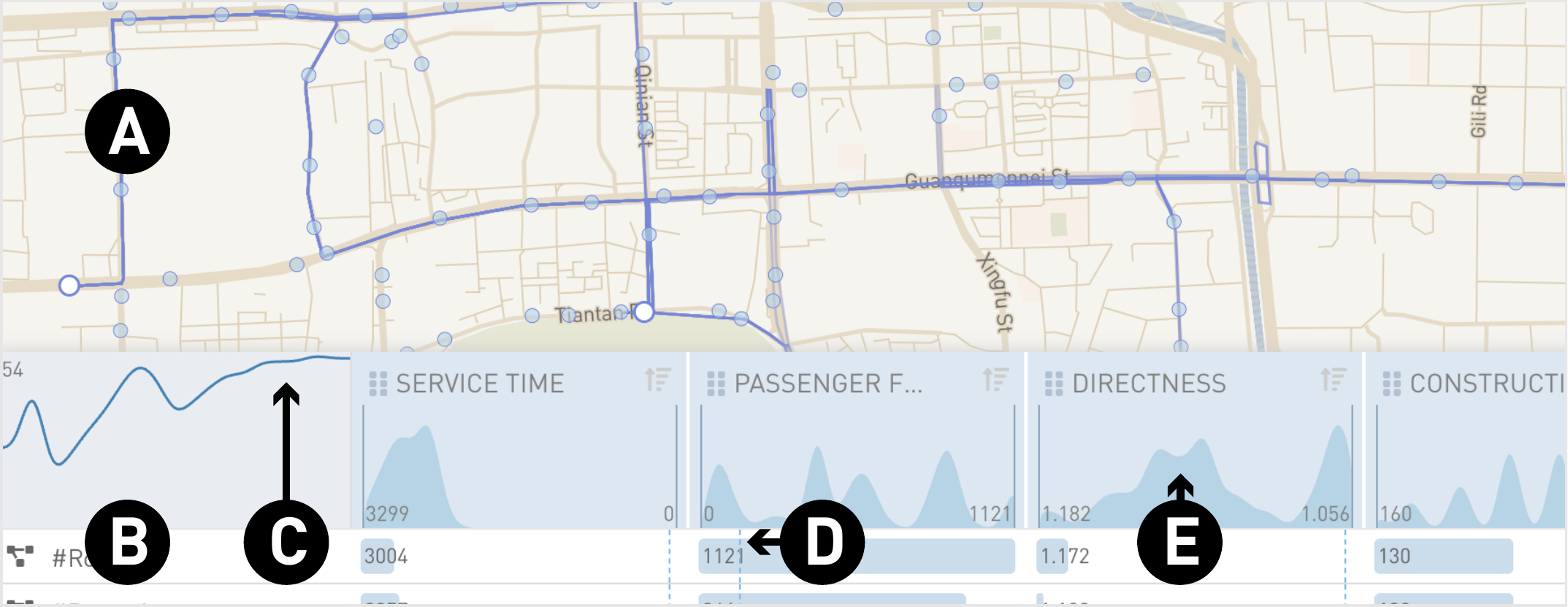}
    \caption{The illustration of the manipulation interface. A) The feasible stops and the generated routes will be displayed on the map. B) The criteria of the routes are depicted in the table. C) An animate line chart shows the number of generated routes. D) The vertical dashed lines indicate the criterion values of the original route. E) The criterion distributions of the obtained routes are shown in the column headers.}
    \label{fig:manipulation}
\end{figure}

After a deficient route has been determined with the exploration interface, the users can obtain replacement routes from the manipulation interface.
This interface allows the users to control the model by specifying model parameters, criterion filters, and anchored stops (M1).
The model results will be streamed and visualized in real-time to help the users determine the quality of the generated routes (M2).

The toolbox at the left of the system (Fig.~\ref{fig:teaser}E) provides fine-grained model controls, including starting/pausing the optimization of the selected route, navigating to the previous/next result set, exiting the manipulation interface, toggling the visibility of the original route, and configuring the parameters for the generation process.
After the generation starts, feasible stops will be detected and shown on the map as blue circles (Fig.~\ref{fig:manipulation}A).
The produced routes are drawn in blue with opacity.
%, where their density indicates the number of overlapping routes.
% After the generation process starts, existing nearby stops will be automatically detected and shown on the map.
The users can anchor the stops with clicks or remove the stops with right clicks.
New stops can also be added by clicking on the map.

The generated routes will be displayed in the route ranking view (Fig.~\ref{fig:manipulation}B) in real-time, allowing the users to evaluate these routes with ranking and filtering.
An animated line plot in the left top of the table (Fig.~\ref{fig:manipulation}C) shows the number of generated routes.
The vertical dashed lines in the table (Fig.~\ref{fig:manipulation}D) indicate the criterion values of the original route.
The criterion distributions in the column headers (Fig.~\ref{fig:manipulation}E) not only allow the users to specify the ranges of criteria but also deliver a quality overview of the generated routes where the users can judiciously determine the termination of the generation process.

\subsection{Evaluation Interface}
\begin{figure}[!tb]
    \centering
    \includegraphics[width=0.4\textwidth]{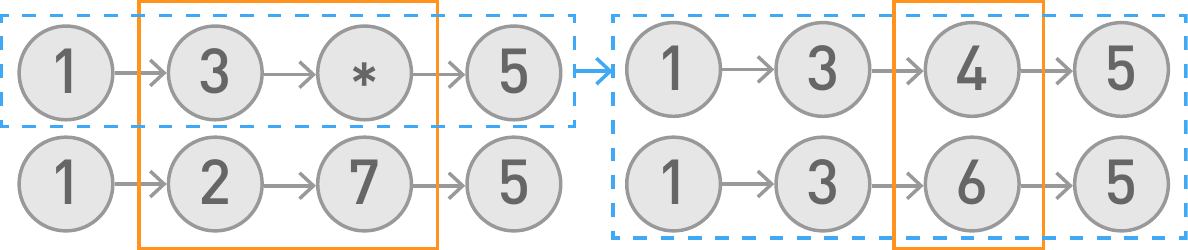}
    \caption{An example of conflicts. The conflicted parts are marked with orange boxes. After aggregation, users have two conflicted choices on the left: \texttt{1-3-*-5} and \texttt{1-2-7-5}. The former choice leads to another two conflicted choices on the right, \texttt{1-3-4-5} and \texttt{1-3-6-5}.}
    \label{fig:conflict}
\end{figure}

To help the users evaluate hundreds of alternative routes and identify the most optimal one, we propose an interactive \textit{conflict resolution} strategy to facilitate efficient and reliable progressive decision-making among these routes.
The strategy is twofold.
First, the topological similarities and differences among the routes are determined with a route clustering method, and the differences between the routes when they go through different streets are grouped into \textit{conflicts}.
% For example, two five-stop routes, \texttt{1-2-5-7-6} and \texttt{1-4-5-8-6}, comprise two conflicts, namely, one at the second stop and another at the fourth stop.
After the conflicts have been extracted, the users can interactively \textit{resolve} these conflicts by iteratively choosing a cluster of routes that share a similar topology and eventually arrive at the most optimal route.
To aid the users in such progressive decision-making processes, the routes and their topologies are depicted with \textit{conflict markers} on the map (L1), and the criteria of the available choices are visualized in the ranking view to facilitate the analysis of route performance (L2).
% For example, to resolve the conflict at the second stop in the two-route example, the users can either choose the stop 2 or 4.

% For example, three four-stop routes, \textit{1-3-4-5}, \textit{1-3-6-5}, and \textit{1-2-7-5}, comprise a nested conflict between the second and third stops.
% If the users choose to resolve this conflict with \textit{3-*} (the asterisk matches any stop), another conflict at the third stop will be presented with the options including the stop 4 or 6.

\subsubsection{Detecting Conflicts}

\begin{algorithm}[!tb]
    \begin{algorithmic}[1]
        \Statex\textbf{Input:} A set of routes $R$, a set of criterion functions $C$ and their weights $w$, and the desired number of choices $\beta$.
        \Statex\textbf{Output:} A set of route clusters $G$.
        \vspace{1mm}

        % \Procedure{MatchRoutes}{$R,g$}
        %     \State\Return $\{r_j\in R|g\subseteq r_j\}$
        % \EndProcedure
        \Procedure{ClusterRoutes}{$R,C,w,\beta$}
            \State $G\gets R$
            \While{$|G|>\beta$} \label{alg:clusterroutes:stopcriterion}
                \State $P \gets \argmax_{ (g_u, g_v) \in G^2 } | g_u \cap g_v |$ \label{alg:clusterroutes:intersect}
                \State $G' \gets \{ g_u \cap g_v | (g_u, g_v) \in P \land \{ g_i \in G | g_u \cap g_v \nsubseteq g_i \} \neq \varnothing\}$ \label{alg:clusterroutes:test}
                \State \algorithmicif\ $G' = \varnothing$\ \textbf{break} \label{alg:clusterroutes:notpossible}
                \State $g_m\gets \argmin_{g\in G'}\Call{StdDev}{\{\sum_{c_k\in C}w_{c_k}c_k(r_i)|r_i\in\{r_j\in R|g\subseteq r_j\}\}}$ \label{alg:clusterroutes:stddev}
                \State $G \gets \{ g \in G | g_m \nsubseteq g \} \cup \{ g_m \}$ \label{alg:clusterroutes:merge}
                % \State $P\gets \argmax_{(g_u,g_v)\in P'}\Call{CriterionDist}{R,C,g_u,g_v}$
                % \State Randomly select a pair of $(g_u,g_v)\in P$ with the smallest criterion difference $\sum_{c_k\in C}|\overline{\sum_{g_u\subseteq r_i\in R}c_k(r_i)}-\overline{\sum_{g_v\subseteq r_j\in R}c_k(r_j)}|$
                % % Find $(r_u, r_v)\in R^2$ which maximizing $|r_u\cap r_v|$ and minimizing $\sum_{c_k\in C}|c_k(r_i)-c_k(r_j)|$
                % \State $N\gets \{g_i\in G|g_u\cap g_v\subseteq g_i\}$
                % \lIf{$N=G$}{
                %     \State \textbf{break}
                % }
                % \State $G\gets G\setminus\{g_u, g_v\}\cup\{g_u\cap g_v\}$
            \EndWhile
            \State\Return $G$
        \EndProcedure
    \end{algorithmic}
    \caption{The route clustering algorithm attempts to extract no more than $\beta$ clusters from the given routes based on their similarities.}
    \label{alg:clusterroutes}
\end{algorithm}

The concept of conflicts assists the users in resolving the topological differences between the routes by comparing and selecting the candidate routes they prefer.
However, excessive routes may lead to choice overloading~\cite{chernev2015choice} in such a complex decision-making scenario, prohibiting the users from identifying optimal routes.
Therefore, we propose to group the routes into several clusters first and then detect the conflicts among the route clusters, such that the choices to resolve each conflict will be no more than these clusters.
For example, three four-stop routes, \texttt{1-3-4-5}, \texttt{1-3-6-5}, and \texttt{1-2-7-5}, can be grouped into two clusters matching the route patterns \texttt{1-3-*-5} and \texttt{1-2-7-5}, respectively, when the number of choices is limited to two.
Hence, a conflict occurs at the second and third stops as shown in Fig.~\ref{fig:conflict}.
If the users choose to resolve this conflict with \texttt{3-*}, another conflict will be detected on the third stop; otherwise, the optimal route \texttt{1-2-7-5} is decided.

The route clustering algorithm is detailed in Alg.~\ref{alg:clusterroutes}.
Initially, we consider each route as a route cluster comprising only one route.
First, we find the pairs of route clusters that share the most stops (cf. line~\ref{alg:clusterroutes:intersect}).
Then, for each pair of clusters, we compute the standard deviation of weighted criterion sums for the routes in a new cluster obtained by merging clusters whose topologies are similar to the common part of these two clusters (cf. line~\ref{alg:clusterroutes:stddev}).
Subsequently, the pair with the minimum standard deviation will be merged (cf. line~\ref{alg:clusterroutes:merge}) if the merge does not reduce the number of choices to one (cf. line~\ref{alg:clusterroutes:test}).
The algorithm repeats this procedure until such a merge is impossible (cf. line~\ref{alg:clusterroutes:notpossible}) or the number of clusters is under the given threshold $\beta$ (cf. line~\ref{alg:clusterroutes:stopcriterion}).

After the routes have been clustered, the conflicts can be detected by aligning all clusters in $G$ by stop and determining the consecutive stops that are not shared among the clusters.

% comprise a nested conflict (the route cluster \texttt{1-3-*-5} vs. \texttt{1-2-7-5}) at the second and third stops.
% If the users choose to

% Based on the detected conflicts, the users can compare among the available routes

% To identify the conflicts and the choices that resolve these conflicts, we first need to extract route clusters

% Moreover, conflicts can be nested.
% For example, three four-stop routes, \texttt{1-3-4-5}, \texttt{1-3-6-5}, and \texttt{1-2-7-5}, comprise a nested conflict (the stop sequences \texttt{3-4} and \texttt{3-6} vs. \texttt{2-7}) between the second and third stops, where \texttt{3-4} and \texttt{3-6} comprise a sub-conflict (the stop \texttt{4} vs. \texttt{6}) at the second stop.

% propose a route clustering algorithm (Fig.~\ref{alg:clusterroutes}) that attempts to generate no more than $\beta$ clusters from a given set of routes.

\subsubsection{Resolving Conflicts}

\begin{figure}[!tb]
    \centering
    \includegraphics[width=0.48\textwidth]{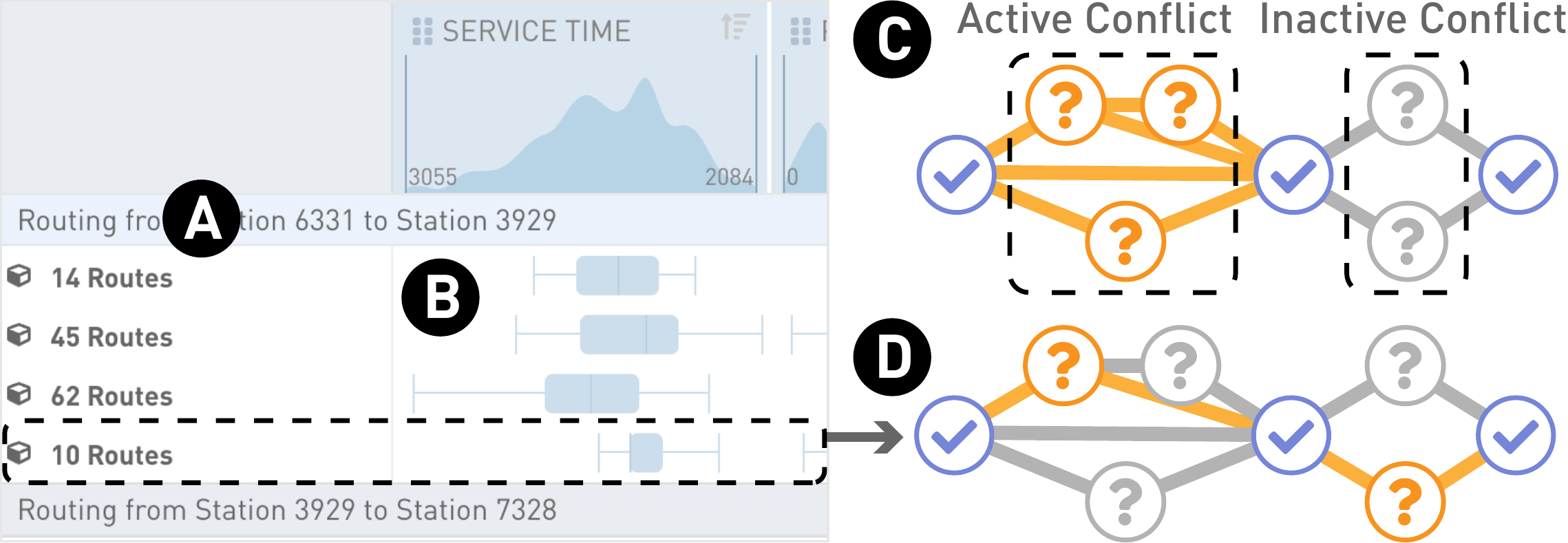}
    \caption{The illustration of the evaluation interface. A) Each collapsible group in the ranking view represents a detected conflict, and each row in the groups is a route cluster. B) The criterion distributions of a route cluster are visualized with boxplots. C) The topology of the route clusters is depicted with a node-link diagram, where the state of each shared stop is indicated with a conflict marker. D) Each row in the table corresponds to a path from the origin stop to the destination stop.}
    \label{fig:evaluation}
\end{figure}

Each detected conflict is shown in the ranking view as a collapsible group (Fig.~\ref{fig:evaluation}A).
Only one conflict is active for resolving at a time.
The users can toggle between the conflicts by clicking on the groups.
The rows in each group represent route clusters, where each of them corresponds to a path from the origin stop to the destination stop, as illustrated in Fig.~\ref{fig:evaluation}D.
The users can hover on a row to see the routes in the cluster on the map and click on a row to resolve the conflict with the cluster.
When a route cluster has more than one route, its criterion distributions are visualized with boxplots (Fig.~\ref{fig:evaluation}B); otherwise, the cluster is displayed similar to a normal route, showing criteria with bars to suggest that this cluster will be the final choice.
% Similarly, the criterion values of the original route will be indicated with a vertical dashed line in each column.

The topology of the route clusters is visualized on the map with a node-link diagram (Fig.~\ref{fig:evaluation}C).
A conflict marker is placed on every stop in the topology showing the conflict status associated with this stop:
1) \textit{resolved} (blue check marks): this stop is shared by all remaining routes;
2) \textit{active} (orange question marks): this stop is shared in a cluster and belongs to the conflict that is currently being examined in the ranking view;
3) \textit{pending} (gray question marks): this stop is shared in a cluster that is waiting to be examined.
The users can obtain the routes that a stop belongs to by hovering on that stop.

% To visualize the predicted passenger flows of the chosen route, the route matrix view shows a flow matrix comprising only the resolved stops.
% Instead of visualizing the trip records, the flow matrix shows the historical passenger flows among the stops regardless of transfers.
% Moreover, if there is a conflict between two consecutive resolved stops, a dashed line will be inserted into the matrix, indicating that the matrix will be expanded at this position.

% \subsubsection{Partial Route Matrix View}

% \subsection{Interactions}

\section{Evaluation}

In this section, we present two usage scenarios and an expert interview to illustrate the effectiveness of the proposed system.

\subsection{Usage Scenarios}

The following two usage scenarios were conducted to analyze and improve the performance of the Beijing bus network in 2013.
The dataset we used comprises 11,414 stops, 653 routes, and 668,346 sampled trip records spanning across two months.
We follow Jim, a transportation system analyst, to see how our system can be used in identifying deficient routes and improving these routes interactively.

\subsubsection{Deficient Route Identification}

\begin{figure}[!tb]
    \centering
    \includegraphics[width=0.48\textwidth]{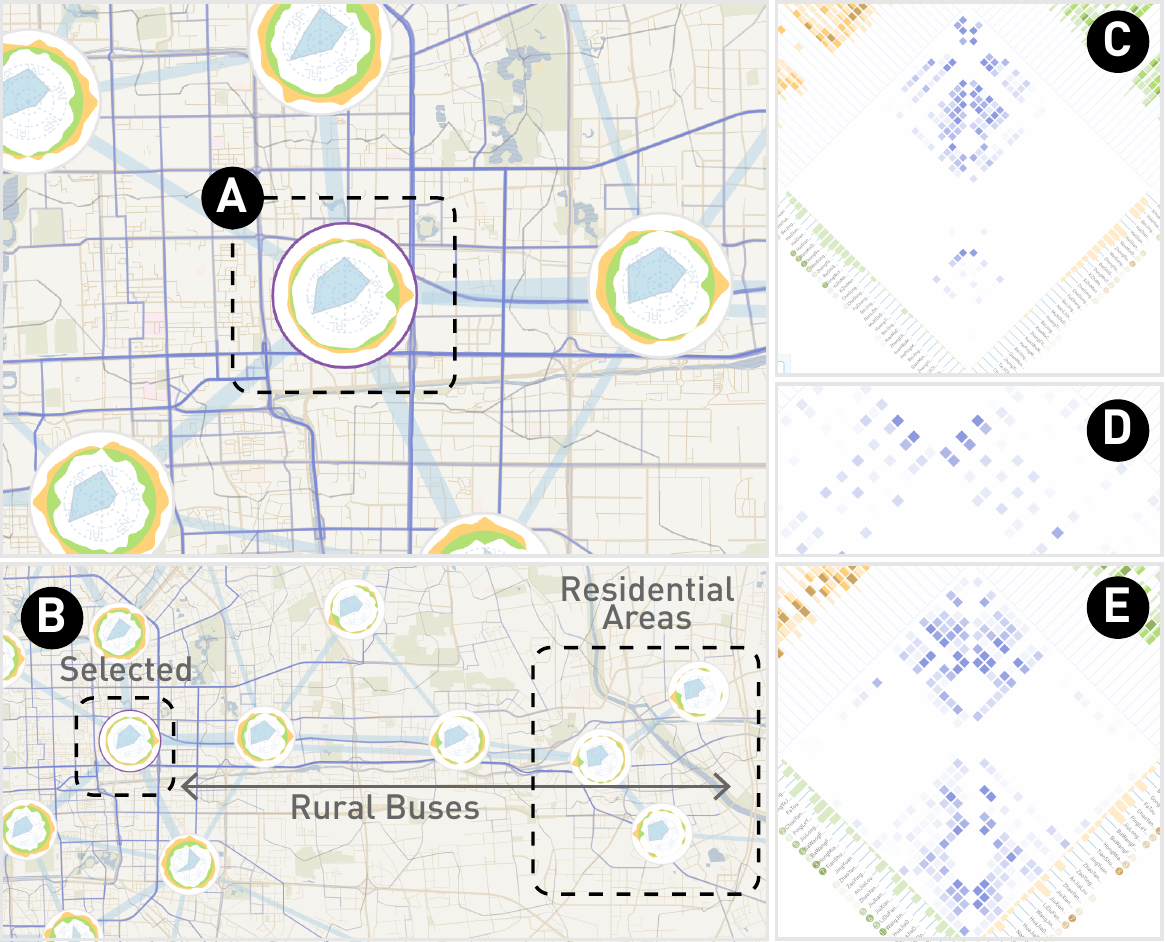}
    \caption{A) Jim was interested in this glyph because of its heavily-biased passenger flows. B) Many rural buses were originated from the selected zone, connecting this zone, a business district, with residential areas. C, D, E) The matrix patterns that indicate that the second half of the route is not functional (C), indicate that the stops are placed too close to each other (D), and suggest that the route can be split (E).}
    \label{fig:case1}
\end{figure}

After Jim loaded the dataset into the system, he saw the map was divided into several transportation zones, whose performance was summarized with the zone glyphs.
The radar charts in the glyphs (Fig.~\ref{fig:teaser}B) revealed that bus routes in the northern part of the city carried more passengers than those in the southern part (higher average load and passenger volume).
He immediately noticed that one glyph (Fig.~\ref{fig:case1}A) was particularly different from others due to its imbalanced passenger flows biased heavily towards the east (outside) of the city.
Hence, he selected this glyph to further analyze all routes that passed this zone.
By judging from the distribution of the routes (Fig.~\ref{fig:case1}B), Jim found that the imbalanced passenger flows were because a) many rural buses were originated from this zone and headed to the east, and b) this zone mainly served as a business district, and many of its employees commuted daily between this zone and the eastern residential districts.

To find potentially deficient routes, Jim applied several filters in the ranking view, including \textit{passenger flows} $\ge$ 10,000 (eliminating the routes with poor data quality), \textit{route length} $\le$ 35 kilometers (eliminating long-distance rural routes), and \textit{service cost} $\ge$ 2,000 (seeking the routes with high service costs).
Nine routes that matched the criteria remained in the ranking view (Fig.~\ref{fig:teaser}F).
Jim inspected them individually and found the following four interesting patterns:

\textbf{Route \#209.}
The ranking view revealed that this route had the lowest average load.
The matrix view (Fig.~\ref{fig:case1}C) further confirmed that the second half of this route was almost empty and non-functional due to low passenger flows.
% Nevertheless, this part of the route still served as short-distance transit between Beijing West railway station and its neighboring stops.
Hence, shortening such a route or removing a few intermediate stops might improve the performance of the bus network.

\textbf{Route \#675, \#701, \#715.}
The matrix view of these routes showed an interleaved pattern (Fig.~\ref{fig:case1}D): the stops adjacent to those with passenger flows had few passenger flows.
Jim determined that this was probably because the stops were placed too close to each other, and potential passengers were absorbed by the nearest stops.
Based on this pattern, these routes could be further optimized by removing a few stops to reduce the service cost and passenger waiting time.

\textbf{Route \#677.}
From the matrix view (Fig.~\ref{fig:case1}E), Jim determined that this route might be split into two because passenger flows in the first and second halves of the routes were independent.
Such a pattern could also guide the scheduling of buses on this route to meet the demand in the rush hours, such as dispatching more buses to run the first half.

\textbf{Route \#683.}
%\marginpar{\todo{add figure references}}
In the matrix view, the hourly temporal distribution of check-out records (Fig.~\ref{fig:teaser}C) clearly distinguished the stops located in the business (passengers checking out around 9 a.m.) or residential (passengers checking out in the evening) areas, which indicated that this route mainly served as a commuting route.
This pattern was also confirmed by the records aggregated based on weekdays.
Furthermore, Jim found this route was particularly different from other ones because of the high transfer volume at the origin stop (Fig.~\ref{fig:teaser}G).
By expanding the transfer route list and examining the pie charts before the routes, Jim found that most passengers transferred to Route \#696.
He then continued to obtain the matrix of Route \#696 (Fig.~\ref{fig:teaser}H), where he found that many passengers boarding this route at the transfer stop were heading towards the city west (from the green stop to the orange stop in Fig.~\ref{fig:teaser}A).
They chose to transfer to Route \#696 because Route \#683 did not extend into the far west region of the city, where the new residential areas were located.
To improve the passengers' experience, Jim suggested that more stops in the west can be added to Route \#683, such that the passengers would not have to make any transfers.

\subsubsection{Interactive Route Replacement}

\begin{figure}[!tb]
    \centering
    \includegraphics[width=0.48\textwidth]{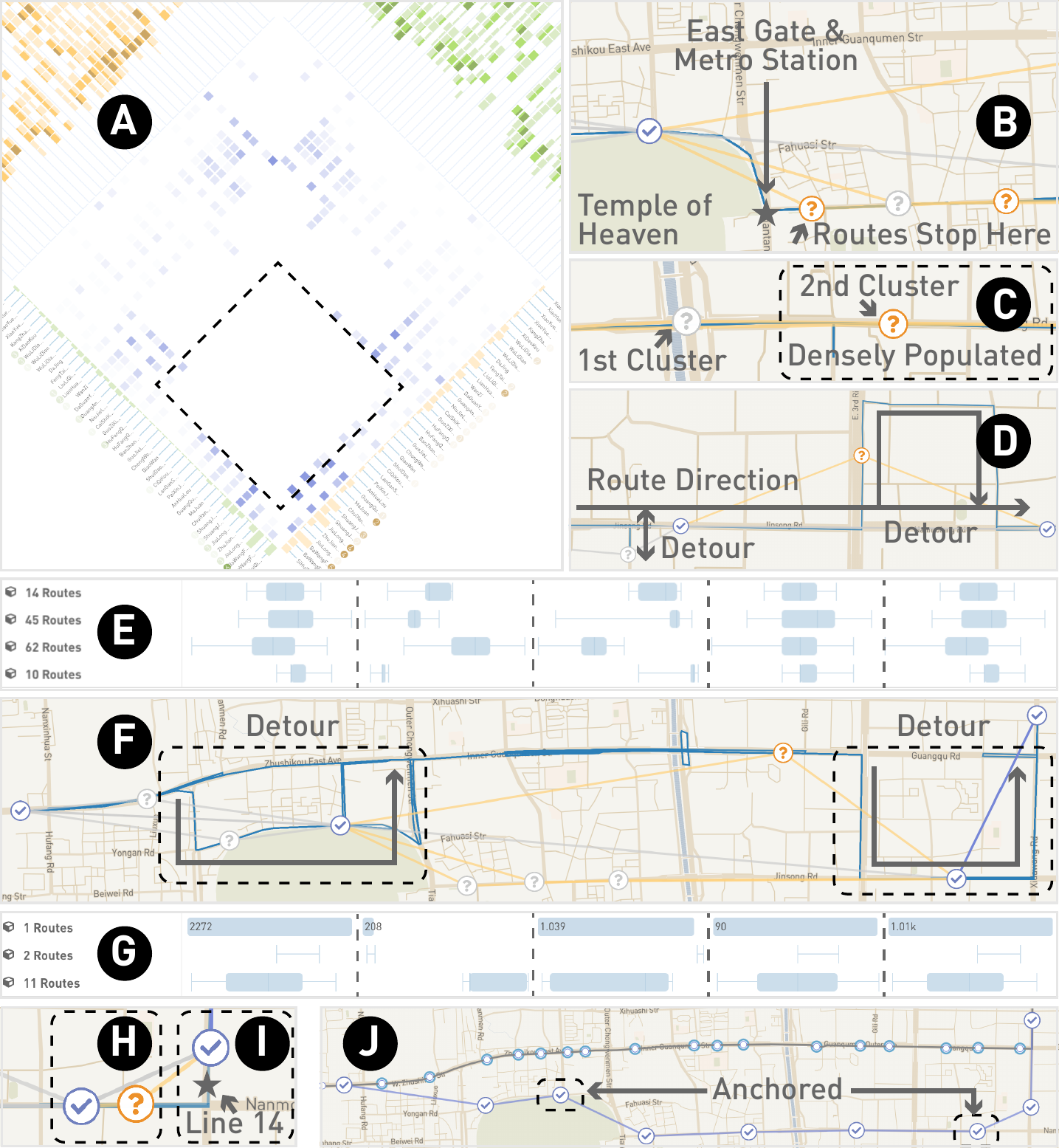}
    \caption{A) The second half of Route \#715 was optimized due to low passenger flows. B) The route cluster stops near the east gate of the Temple and a metro station. C) The second cluster is preferred because it stops in a densely populated area. D, F) The route clusters take multiple detours. E, G) The route clusters generated in step 1 (E) and 2 (G). The criteria from left to right: service time, passenger flow, directness, construction cost, and service cost. H) The stops are too close. I) This stop is located near a metro station. J) The final route (stops are marked with checkmarks) obtained from the decision-making process.}
    \label{fig:case2}
\end{figure}

In the ranking view (Fig.~\ref{fig:teaser}F), Jim ranked the routes with the passenger volume and service cost criteria combined and discovered that Route \#715 performed worst due to its low passenger volumes.
By checking the matrix view (Fig.~\ref{fig:case2}A), Jim found that the second half of this route was not functional because this part of the route largely overlapped with a metro line.
Hence, after observing the map, Jim decided to optimize this route by rerouting it towards the south, such that the route would pass through a tourist attraction (Temple of Heaven) and two hospitals.

Jim selected the part of the route to be optimized and anchored two stops on the map (Fig.~\ref{fig:case2}J), one at the north gate of the Temple and another one near the hospitals.
He then started the model and observed the route generation progress in the ranking view.
The generation was terminated as the distributions of criteria stabilized, leaving 132 alternative routes in the ranking view.
We show how Jim decided the most optimal route progressively with the following four steps.

\textbf{Step 1.}
Jim obtained two conflicts and four route clusters from the ranking view (Fig.~\ref{fig:case2}E).
He observed from the boxplots that the third cluster had the largest passenger flows.
This cluster mainly followed the original route and satisfied the stop anchoring constraint by taking two detours (Fig.~\ref{fig:case2}F), resulting in particularly low route directness.
Therefore, Jim inspected the first two clusters, which had the second-largest passenger flows, and found these clusters shared similar characteristics in terms of performance criteria.
He then examined these two clusters on the map and identified that the routes in the first cluster would stop at the east gate of the Temple and a nearby metro station (Fig.~\ref{fig:case2}B), gaining potential passenger flows.
Hence, the first cluster was selected.

\textbf{Step 2.}
After the selection, the ranking view (Fig.~\ref{fig:case2}G) presented another two conflicts with three clusters, one of which comprised only a route.
The third cluster had the most passenger flows and provided more diverse choices compared with the first and second because the routes in this cluster were connected with the metro line 14 via the stop illustrated in Fig.~\ref{fig:case2}I.
Therefore, Jim proceeded with this cluster.

\textbf{Step 3.}
As the decision-making process continued, one conflict with two clusters was depicted in the ranking view.
The difference between these two clusters shown on the map (Fig.~\ref{fig:case2}C) with the conflict markers was that the routes in the first cluster stopped near a bridge and those in the second stopped at dense residential areas.
Therefore, the second cluster was chosen due to its potential in attracting more passengers.

\textbf{Step 4.}
Thereafter, Jim was presented with another three conflicts with three clusters, the first and second of which comprised only a route.
For the third cluster which had the most passenger flows, the first and second conflicts indicated that the flows were obtained by taking multiple detours (Fig.~\ref{fig:case2}D), which greatly impacted the bus riders' experience.
For the second route, Jim noticed that, in the third conflict, two stops in this route were too close to each other (Fig.~\ref{fig:case2}H), which might lead to the interleaved flow pattern.
Hence, Jim concluded that the first route (the blue route in Fig.~\ref{fig:case2}J) could be the most optimal one, given its balanced topology and performance criteria.

\subsection{Expert Interview}

To collect feedback from experts, we conducted structured interviews individually with four domain experts.
First, we introduced the basic functionalities of the system.
Then, we demonstrated the usage scenarios to show how the system can be used in detecting and improving deficient routes.
Thereafter, we asked the experts to follow the think-aloud protocol and try to explore the bus network and improve a few routes with the system.
Their comments during the process were recorded and analyzed.
Finally, we requested them to give additional feedback on the usability and effectiveness of the system.

EA spoke highly of the transportation zones and zone glyphs.
``\textit{An overview of the bus network is clearly given,}'' commented EA. %, ``\textit{and I especially like the idea that the glyphs encoding the spatial information of passenger flows with the rings.}''
EB shared similar opinions with EA, saying that transportation zones showed a clear structure of the bus network and also facilitated better route filtering.
Moreover, EB and EC were both impressed by the matrix view.
EB commented that ``\textit{A lot of insights can be obtained from such a view.}''
%EC also commented that ``\textit{I have never analyzed bus routes with this type of visualization before. It establishes a new way to evaluate route performance.}''
Furthermore, EA, EB, and ED praised the conflict-resolving strategy in route decision-making, commenting that such a strategy was ``\textit{intuitive}'', ``\textit{easy to use}'', and ``\textit{engaging}''.

While the experts confirmed the design of our system had met with their analytical requirements, they also gave feedback on how to improve the usability of the system.
EA requested that the system shall allow the filters in the ranking view to be reset.
EC suggested that an animated line chart showing the number of the generated routes could further help the users control the model.
ED recommended adding support for additional geometry file types to enhance the compatibility with other software.
The system had been improved accordingly.

To further verify the effectiveness of the proposed system, we have planned to continuously collaborate with the experts after the expert interview to deploy the system for production use and test it in real-world scenarios, which will allow us to fully evaluate the effectiveness of the system with practical use cases in the field.

\section{Discussion}

In this section, we discuss the implications, lessons learned, limitations, and future work of the proposed system.

\textbf{Implications.}
%This study introduces BNVA, a novel visual analytics system for analyzing and improving bus route networks.
%To the best of our knowledge,
This study is the first step towards the model-assisted visual exploration and planning of bus routes.
In terms of \textit{techniques}, BNVA comprises novel visualizations and decision-making strategies to facilitate the improvement of bus routes.
In terms of \textit{evaluation}, the usage scenarios indicated several abnormal route patterns that may provide insights for the bus network planners and guide the design of effective bus routes in the future.
In terms of \textit{applicability}, the progressive decision-making strategy proposed in this study can be applied to many urban planning scenarios, where the overwhelming number of choices will be reduced to help users make decisions judiciously.
% Three interfaces, including the exploration, manipulation, and evaluation interfaces, are developed to facilitate the detection and optimization of deficient routes.
% The exploration interface leverages an overview-to-detail approach that comprises the network, route, and station levels to assist the users in the efficient analysis of bus route performance.
% The manipulation interface establishes the user-in-the-loop route generation processes with the intuitive visual indicators of route quality to help the users effectively monitor and control the model results.
% The evaluation interface aggregates the generated routes based on a route clustering method and facilitates an informed progressive decision-making process with the conflict-resolve strategy.
% By integrating these three interfaces into an iterative workflow, BNVA enables the transportation planning experts to visually assess the performance of bus networks and routes and improve the deficient routes with model-generated alternatives.
% To the best of our knowledge, BNVA is the first step towards the model-assisted visual exploration and planning of bus routes.

\textbf{Lessons learned.}
We present two design lessons we have learned while developing BNVA.
First, hierarchical exploration is mandatory for complex data like bus networks.
The complexity of bus networks not only lies in the horizontal (network size) and vertical (zones, routes, and stations) scales of the networks but also emerges from the association between the networks and other data sources, such as passenger flows.
%To allow the users to comprehend the performance of bus networks from different perspectives,
In this study, we employed a hierarchical exploration approach to dissect a network with multiple coordinated views from the overview to the details and provide visual hints for the users to drill down in the dataset and find interesting patterns.
Second, progressive decision-making techniques can be useful in reducing the difficulties in making complex decisions.
Excessive choices may overwhelm the cognitive capability of users.
In the evaluation interface, we propose a progressive decision-making strategy that assists the users in comparing and evaluating the generated routes through choice aggregation.
Users can eliminate a large number of choices efficiently because the choices that share similar characteristics are aggregated.
We observed that the experts rapidly adapted to such a strategy and gave positive feedback.

\textbf{Limitations and future work.}
Two limitations are observed in this study.
First, the route generation model does not consider route shape requirements.
%Many cities have circular bus routes for several reasons, including reducing the number of bus pools.
Although circular routes can be planned by adding multiple anchored stops, direct support of specifying route shapes will facilitate design flexibility.
This functionality can be achieved by integrating route shape optimization, which we will leave as a part of the future work.
The second limitation involves the scope of our study.
Planning an optimal and practical bus route not only requires optimized spatial structure but also a well-designed timetable and crew assignment.
However, this study primarily focuses on the spatial optimization of the existing routes.
%, allowing the users to export the optimized routes to other dedicated GIS software for subsequent resource scheduling.
In the future, we plan to implement other planning stages to establish a complete pipeline of bus network design.

There are some other potential future directions.
BNVA is highly extensible in supporting customized quantifiable criteria.
Additional data sources, such as traffic data and other transportation networks, can be incorporated to improve the performance of alternative route extraction and evaluation.
Some machine learning approaches~\cite{DBLP:journals/chinaf/Zhou19} may improve the performance of route generation.
Furthermore, integrating choice recommendations in selecting the optimal routes will considerably improve the efficiency of decision-making.
Such integration will be studied in the future to facilitate spatial decision-making.

\section{Conclusion}

This study presents BNVA, a novel visual analytics system that helps experts visually explore bus networks, identify deficient routes, obtain candidate routes, and determine the best alternative.
The exploration, manipulation, and evaluation interfaces are designed to tackle three identified challenges, namely, the in-depth analysis of complex bus route networks, the interactive generation of improved route candidates, and the effective evaluation of alternative bus routes.
The proposed system has been evaluated with usage scenarios and expert interviews based on real-world data.
In the future, we will implement additional route generation criteria and expand the capability of BNVA by integrating scheduling and resource management functionalities.

\acknowledgments{
  The work was supported by  NSFC-Zhejiang Joint Fund for the Integration of Industrialization and Informatization (U1609217), National Key R\&D Program of China (2018YFB1004300), NSFC (61761136020), Zhejiang Provincial Natural Science Foundation (LR18F020001) and the 100 Talents Program of Zhejiang University. }

\bibliographystyle{abbrv}

\bibliography{paper}
\end{document}